\documentclass[preprint]{aastex}

\def\DD{{{I\negthinspace\!D}}}
\def\LL{{{I\negthinspace\!L}}}

\begin{document}

\title{\bf The Radiative Stress Tensor}
\author{Xinzhong Chen and Edward A. Spiegel}
\affil{Astronomy Department, Columbia University, New York, NY 10027}
\email{xinzhong@astro.columbia.edu}

\begin{abstract}

We use the transfer equation in relativistic form to develop an
expansion of the one-photon distribution for a medium with constant
photon mean free path, $\epsilon$.  On carrying out appropriate
integrations and manipulations, we convert this expansion into one for the
frequency-integrated intensity.  We regroup the terms of the
intensity expansion according to both the power of $\epsilon$ and the
angular structure of the various terms and then carry out angle
integrations to obtain the expansions for the components of the stress
energy tensor: the radiative energy density, the radiative flux and the
pressure tensor.  In leading order, we recover Thomas' (1930)
results for the viscosity tensor and his expression for the viscosity
coefficient, which are correct for short mean free paths.  As had been
done earlier for the radiative heat equation \citep{unn66}, we keep at
each order in the expansion a dominant portion, but this time one with a
richer angular structure.  Then, after some rearrangement of the various
summations in the expressions for the moments, we replace the sum of the
calculated higher order terms by a Pad\'e approximant, or rational
approximation \citep{bak75}, to provide an improved closure approximation
for the radiative stress tensor.  The resulting radiative viscosity tensor
may be expressed either as a simple integral operator acting on the Thomas
stress tensor or as the solution of an inhomogenous, linear partial
differential equation.  The expression obtained for the radiative
viscosity tensor applies for media with long, as well as short, photon
mean free paths.  We also develop results applicable for relatively smooth
flows by using the form of the Thomas stress tensor with generalized
transport coefficients derived by the application of a suitable operator
to the bare Thomas coefficients.

\end{abstract}


\section{Introduction}

The presence of a radiation field will subject an otherwise perfect
fluid to dissipation.  In the limit of short photon mean free paths,
dissipation is expressed as diffusion of momentum and energy and is
regulated by the transport coefficients of viscosity and conductivity.
These were obtained by \citet{tho30} who started from the
relativistic transfer equation and developed the solutions in an
expansion in photon mean free path.  Since Thomas retained only terms of
first order, his was a diffusion theory whose results have been extended
to include the effects of general relativity \citep{ham71} Thomson
scattering \citep{mas71,hsi76} and Compton scattering \citep{mas81}.

Unfortunately, a diffusion approximation does not provide an adequate
description of the effects of radiation on the medium and its dynamics
when vigorous radiative fluid dynamics produces large fluctuations in
physical conditions.  In such conditions, transparent regions may form
and the photon mean free path can become relatively long so that better
approximations are needed.  One approach is to continue using the
equations for the first two moments of the radiative intensity, energy
density and flux, and closing the system by providing an improved
approximation for the pressure tensor.  A step in this direction is to
use the Eddington factor methods \citep{mih84} as has been done by a
sequence of authors \citep{cas72,fua87,min78,dom95,dom96,dom97}.  As
Castor (1972; see also \citep{mih83}) has reported, the scheme seems
to be usable for long photon mean free paths, but it does call for the
introduction of additional information.  Typically the variable Eddington
factor is to be found from numerical solutions of the transfer equation in
related conditions.  It is therefore desirable to seek a self-contained
approximation for the pressure tensor and we shall provide one here.

We shall here use an approach that originates, in part, with a method
used for closure in the study of thermal effects of radiation
\citep{unn66}.  In the treatment of the thermal effects alone, the
Eddington approximation
gives rather good results.  This is so because the Eddington closure
relation for the radiative stress-energy tensor is the result of
resummation of terms of all orders in an expansion in photon mean free
path \citep{unn66}, and it captures aspects of both optically thin and
thick structures.  When we use the Eddington approximation we get a
radiative heat equation (for a static medium) that is schematically of the
form \citep{unn66}: \begin{equation}
\partial_t S = \kappa_1{\cal D}\,\nabla^2\, S \ , \qquad {\cal D} =
\left(1 - \kappa_2 \nabla^2\right)^{-1} \ , \label{diff} \end{equation}
where $\kappa_1$ and $\kappa_2$ are suitable ``diffusion'' coefficients.
The Laplacian, which is the usual diffusion operator, is here augmented by
the additional smoothing operator ${\cal D}$.  Of course, this is the
bare essential of such a result, whose simplicity depends on a number of
idealizations, most notably that the mean free path of photons is a
constant, and on the neglect of retardation effects.  Nevertheless, ${\cal
D}$ carries much of the qualitative character of the radiative
smoothing process.  If we develop the smoothing operator in equation
(\ref{diff}) for large scales (small $\nabla^2$, qualitatively speaking)
we obtain ${\cal D}\nabla^2 \approx \nabla^2 +  \kappa_2 \nabla^4$.  This
illustrates why we may recover the full expression for ${\cal D}$ from a
Pad\'e approximation based on only the first two terms in the expansion.

On the other hand, for the case of viscous stresses, the Eddington
closure approximation fails to produce a physically acceptable result
\citep{and72} and an improved version is needed.  The aim of the present
paper is to devise a closure for the second order moments of the
radiation field that may be used to deal with both the dynamical
and thermal effects of the radiation field for both optically thick and
thin structures.  Similar techniques have recently been used in kinetic
theory where an improved viscous stress tensor has been obtained by
resumming terms up to only second order in an expansion in mean free path
\citep{ros89, ros93, sle97}.  The resummed forms, when expanded, contain
terms of all orders, as in the earlier radiative thermal study.

Starting from the manifestly covariant form of the transfer equation for a
moving medium, we develop the solution of the transfer equation in
photon mean free path.  Then we group terms within this expansion
according to angular complexity.  This leads to multiple sums and we
rearrange the series so as to be able to compare terms of like character.
From each order, we take the dominant term in power of the mean free path.
Then we choose the leading portion at each order in terms of angular
complexity, in what is really a Galerkin truncation.  The resulting
infinite sum is then converted to finite form by a resummation, or Pad\'e,
technique \citep{bak75} to provide a compact form for the radiative
viscosity tensor.

\section{Basic Formulas}

\subsection{Transfer Equation}

To include the effects of differential motion of a medium on the transfer
of radiation, \citet{tho30} used the transfer equation in
relativistic form. In this way, he readily included effects such as
aberration and Doppler shift and was able to obtain the correct
radiative viscosity where others had failed (Mihalas 1983).  Even if one
is not concerned with relativistic fluid velocities, it is convenient
adopt the relativistic formulation and also to follow \citet{haz59} in
using the covariant form of the equation of
radiative transfer for the case of moving media.  This approach is
especially effective if one makes use of the one-particle distribution
function for photons, $f(p, x)$, where $p$ (or $p^{\mu}$) is the
four-momentum of a photon and $x$ (or $x^\mu$) is its location in
space-time.  According to the way one proceeds, $f$ is a scalar (as here)
or a scalar density.

If we neglect gravitational bending of light and refraction by the
medium, we may write the transfer equation in the general form
\citep{and72}
\begin{equation}
p^\mu \partial_{\mu}{f}  = \tilde\rho (\alpha-\beta f)
\label{trans}
\end{equation}
where $\partial_\mu$ stands for $\partial/\partial x^\mu$ and
summation over any repeated index is understood.  Here
$\alpha$ is the rate of injection of photons with momentum $p$ by
emission and $\beta f$ is the rate at which photons are
removed by absorption; both rates are per unit mass.  The quantity
$\tilde\rho$ is the number density of absorbers times the rest mass per
absorber.

Unlike $f$, the specific intensity, ${\cal I}$, is not a scalar. Its
transformation properties, as given by \citet{tho30}, may be deduced from
the definition
\begin{equation}
{\cal I} = {h^4 \nu^3 f\over c^3} \ ,
\label{int}
\end{equation}
where $h$ is Planck's constant, $c$ is the speed of light and $\nu$
is the frequency.  We shall use units in which $c$ and $h$ are unity.

The transformation properties of the other quantities in the usual
transfer equation are obtained by comparing that equation in the local
rest frame of the matter with equation (\ref{trans}).  For our purposes,
there are two important reference frames, a basic inertial frame (such as
that of a stationary star) and a frame locally comoving with the matter.
The four-velocity of the matter, $u^\mu$, has components $(1,0,0,0)$
in the locally comoving frame.

The invariant
\begin{equation}
\tilde \nu =u^\mu p_\mu    \ .
\label{nu0}
\end{equation}
is the photon frequency in the local rest frame of the matter.  By
comparing the transfer equation in a general frame with that in the
comoving frame, \citet{tho30} found $\beta = \tilde\nu \kappa(\tilde\nu)$,
where $\kappa$ is the opacity, and $\alpha = \tilde\nu^{-2}j(\tilde\nu)$,
where $j$ is the emissivity.  Since $\alpha$, $\beta$ and $\tilde\nu$ are
scalars, we then know the transformation rules for $j$ and $\kappa$ once
we introduce the (relativistic) Doppler relation between general frequency
$\nu$ and $\tilde\nu$.

\subsection{Radiative Stress Tensor}

A basic object of radiative fluid dynamics is the radiative
stress tensor, $T^{\mu\nu}$.  It may be defined as an integral of the
distribution function over momentum space:
\begin{equation}
T^{\mu\nu}=\int p^\mu p^\nu fdP \ ,
\label{stress}
\end{equation}
where $dP$ is the covariant volume element in momentum space. With
$\tilde\nu$ as the radial coordinate in spherical polar coordinates in
momentum space the element of volume in momentum space is
$dP = \tilde\nu d\tilde\nu d\Omega$, where $d\Omega$ is the element of
solid angle \citep{lan84}. We may then factor the frequency from
$p^\mu$ and define the null vector
\begin{equation} n^{\mu} =  p^{\mu}/\tilde\nu \label{nmu} \; ,
\end{equation} so that the stress tensor is \begin{equation}
T^{\mu\nu}= \int\, d\Omega n^\mu \, n^\nu \, I \end{equation}
where we have introduced the frequency-integrated specific intensity
\begin{equation}
I=\int_0^{\infty}\tilde\nu^3 f\, d\tilde\nu \ . \label{IntI}
\end{equation}

The projection operator \begin{equation}
h^{\mu\nu}= \eta^{\mu\nu}-u^\mu u^\nu  \label{hmunu} \end{equation}
 picks out space-like components in the frame of the matter, where
$\eta^{\mu\nu}$ is the Minkowski metric, diag$(1,-1,-1,-1)$.  Then we
may introduce the spacelike unit four-vector, \begin{equation}
l^\mu = h^{\mu}_{\;\nu}\,n^\nu \  , \end{equation}
so that \begin{equation}
n^\mu = u^{\mu}+l^{\mu} \; . \label{lmu} \end{equation}
We see that $l^\mu$ must be perpendicular
to $u^\mu$ and that $l_\mu l^\mu =-1$.
Then with the introduction of (\ref{lmu}) we arrive at a
standard way to write the stress tensor, namely \begin{equation}
T^{\mu\nu}= P^{\mu\nu}+ F^{\mu}u^{\nu}+u^{\mu} F^{\nu}+
E u^\mu u^\nu \
\label{decomp}
\end{equation}
with
\begin{equation}
\label{en}
E= \int I\, d\Omega \ , \ F^\mu = \int  l^{\mu}\, I \, d\Omega \ , \
P^{\mu \nu} = \int l^\mu l^\nu\, I\, d\Omega \ ,
\end{equation}
where these integrals are over all solid angle.

By taking moments of the transfer equation itself, we derive
equations for moments like those in (\ref{decomp}).  This procedure leads
to an infinite hierarchy of moment equations that may be
truncated by introducing a relation amongst the moments,
$E$, $F^\mu$ and $P^{\mu\nu}$.   Such closure relations can be found
by assuming some particular forms for the angular dependence of $I$,
usually expressed in terms of the $l^\mu$ or $n^\mu$ and this requires
angular integrals of their products.  Though we do not follow this
procedure, we do need such integrals and so we give a list of
formulas for the angular integrals in Appendix \ref{angu}.  In the
simplest case, when the isotropic approximation is introduced
for the intensity, we need the integral  \begin{equation}
\int l^{\mu}l^{\nu}d\Omega  =   -\frac{4\pi}{3}h^{\mu\nu}
\label{M2} \end{equation} for the evaluation of $P^{\mu\nu}$
that is used in the Eddington closure approximation \citep{and72}.

\subsection{The Matter and the Total Stress Tensors}

To describe the matter macroscopically, we adopt, with \citet{wei71},
the procedures given by \citet{eck40}, who noted that a vector $V^\mu$
may be written as $V^\mu= (u_\alpha V^\alpha)u^\mu
+ h^\mu_\alpha V^\alpha$, as in equation (\ref{lmu}).  A similar
decomposition is possible for tensors and, in particular, for the
radiative stress tensor, we see that
\begin{equation}
E = u_\alpha u_\beta T^{\alpha \beta}\, , \quad
F^\mu =u_\alpha T^{\alpha \beta}h^{\mu}_\beta\, , \quad
P^{\mu\nu} = h^{\mu}_\alpha h^{\nu}_\beta\ T^{\alpha \beta}\, .
\label{comps} \end{equation}

As in \citet{eck40}'s approach, we may use the form of
(\ref{comps}) to define the three important moments of kinetic theory in
terms of the matter stress tensor. If we denote the stress-energy tensor
of the matter as ${\cal T}^{\mu\nu}$, the mass-energy density, the matter
flux and the pressure tensor are given (as in (\ref{comps})) by
\begin{equation}
{\cal E} = u_\alpha u_\beta {\cal T}^{\alpha \beta}\, , \quad
{\cal F}^\mu = u_\alpha {\cal T}^{\alpha \beta}h^{\mu}_\beta\, , \quad
{\cal P}^{\mu\nu} = h^{\mu}_\alpha h^{\nu}_\beta\ {\cal T}^{\alpha \beta}\, .
\label{matt}\end{equation}
To concentrate on radiative effects, we assume that, in the absence
of interaction with radiation, the matter would be a perfect fluid,
for which ${\cal F}^\mu=0$ and ${\cal P}^{\mu\nu}={\cal P} h^{\mu\nu}$,
where ${\cal P}$ is the partial pressure of the matter. The matter stress
tensor in that case would be
\begin{equation}
{\cal T}^{\mu\nu} =
{\cal E}u^\mu u^\nu - {\cal P} h^{\mu\nu} \ . \label{mattstr}
\end{equation}

The total stress tensor (matter plus radiation) is
\begin{equation}
\Theta^{\mu\nu} =  T^{\mu\nu} + {\cal T}^{\mu\nu}
\label{tot}
\end{equation}
and the equations of motion of the mixture are expressed as
\begin{equation}
\Theta^{\mu\nu}_{\ \ , \nu} = 0\, . \label{eqmo}
\end{equation}

\section{Diffusion Theory}

For a discussion of the extension of Thomas' theory, a brief restatement
of his results is useful for bringing out some of the issues to be faced.
To provide one here, we introduce the source function,
${\cal S}=\alpha/\beta$, and the local photon mean free path,
$\varepsilon=(\tilde\rho\kappa)^{-1}$ where $\kappa$ is generally a
function of $\tilde\nu$ as well as of the state variables of the medium.
The mean free path appears explicitly in the transfer equation
(\ref{trans}), which we rewrite as \begin{equation}
f = {\cal S} - \varepsilon {n^\mu} f_{,\mu}
\label{retrans}
\end{equation}
where $n^\mu$ is defined in (\ref{nmu}).
This way of writing the transfer equation is meant to take advantage
of the smallness of $\varepsilon$, which is assumed in Thomas' theory.
The fact that $\varepsilon$ appears in front of the derivative means
that the perturbation theory based on its smallness will be singular.
Generally speaking, singular perturbations expansions give rise to
(divergent) asymptotic series but these may be made useful by appropriate
(re)summation techniques \citep{ben78} such as we shall use
here.

In first approximation, we have $f\approx {\cal S}$.  If we iterate
on this once in (\ref{retrans}) we get Thomas' approximation,
\begin{equation}
f = {\cal S} - \varepsilon n^{\mu}{\cal S}_{,\mu} \ .
\label{f1a}
\end{equation}

To compute the stress tensor, we first perform the integration over
frequency after multiplying by $\tilde \nu^3$ to obtain an
approximation for the frequency-integrated intensity, \begin{equation}
I = S - n^\mu
\int_0^\infty \tilde \nu^3 \varepsilon {\cal S}_{,\mu}
\; d\tilde \nu \; ,  \label{III}
\end{equation}
where we have introduced the integrated source function
\begin{equation}
S(T)=\int^\infty_0\tilde\nu^3{\cal S}(\tilde\nu,T)d\tilde\nu \ .
\label{intS}
\end{equation}
The integral in formula (\ref{III}) is a simple example of the kind of
integrals that the higher order development entails, so it is worth
discussing its evaluation.

In Thomas theory, $\tilde \nu^3{\cal S}$ is the Planck distribution
and it depends explicitly on $\tilde \nu$ and temperature, $T$, from
which it derives an implicit dependence on position and time.  Therefore
we may write \begin{equation}
{\cal S}_{,\mu} = \tilde \nu_{,\mu} \partial_{\tilde \nu} {\cal S}
+ T_{,\mu} \partial_T {\cal S} \; . \label{deriv} \end{equation}
On recalling that $\tilde\nu =u_\mu p^\mu$ and that $p^\mu$ does not
depend on coordinates, we obtain $\tilde{\nu}_{,\mu} =
p_\nu u^\nu_{, \mu}$. Then, since
$u_\nu u^\nu_{\, ,\mu}=0$, we have that  \begin{equation}
\tilde{\nu}_{,\mu} = \tilde\nu n_\nu u^{\nu}_{\, , \mu}
 \ . \label{commanu} \end{equation}
If, as in local thermodynamic
equilibrium, the source function depends on $\tilde\nu$
and $T$ only through the combination $\tilde \nu/T$, we have
\begin{equation}
\tilde \nu \partial_{\tilde\nu} = - T \partial_T \label{switch}
\end{equation}
and so (\ref{deriv}) can be rewritten as \begin{equation}
{\cal S}_{,\mu} = \left(T_{,\mu}\, - T n^\rho u_{\rho,\mu}\right)
\partial_T {\cal S}\; . \label{partialt} \end{equation}
The integral in (\ref{III}) then becomes \begin{equation}
\int_0^\infty \tilde \nu^3 \varepsilon {\cal S}_{,\mu} \; d\tilde \nu
= (T_{,\mu} - T n^\nu u_{\nu,\mu})
\int_0^\infty \tilde \nu^3 \varepsilon \partial_T{\cal S}
\; d\tilde \nu \; . \end{equation}

If we pull the $T$ derivative out of the integral, we get an extra
term involving the $T$ derivative of $\varepsilon$.  On the other hand,
we can simply introduce a weighted frequency integral through the relation
\begin{equation}
\bar \varepsilon S_{,T} = \int^\infty_0\tilde\nu^3{\cal S}_{,T}(\tilde\nu,T)
\varepsilon d\tilde\nu\ . \label{mean} \end{equation}
The quantity $\bar \varepsilon$ is a mean of the inverse of the absorption
coefficient with respect to ${\cal S}_{,T}$ in the spirit of the
Rosseland mean absorption coefficient.  Then we find that, for
$TS_{,T} = 4S$, \begin{equation}
I = S - \bar \varepsilon n^\mu S_{,\mu} + 4\bar \varepsilon n^{\mu} n^\nu S
u_{\mu,\nu} \; .
\label{intint1} \end{equation}

In the higher order theory to be discussed below, we can use similar
tricks by introducing a variety of mean absorption coefficients, as is
done sometimes in the theory of stellar atmospheres.  More often than
not, all these are assumed to be equal in the final analysis.  To avoid
such purely formal maneuvers, we shall assume in the presentation of
higher order theory that the medium is gray.

Then, to this approximation, partial integrations such as we have
just described give us the known forms of moments of the
intensity in the Thomas approximation:
\begin{equation}
E = 4\pi S - 4\pi \bar \varepsilon \left
[u^\alpha S_{,\alpha}\, +\frac{4}{3} S u^\alpha_{,\alpha}\right],
\label{enden} \end{equation}
\begin{equation} F^\mu = \frac{4\pi}{3} \bar \varepsilon h^{\mu\alpha}
\left[S_{,\alpha}-4S u^\beta u_{\alpha,\beta}\right], \label{flu}
\end{equation}
\begin{eqnarray}
P^{\mu\nu} = -\frac{1}{3} h^{\mu\nu}E+\frac{16\pi}{15}\bar \varepsilon
S u_{\rho,\sigma}
\tau^{\mu\nu\rho\sigma} , \label{press2}
\end{eqnarray}
where we have used the relevant formulae from Appendix \ref{angu} and
have introduced \begin{equation}
\tau^{\mu\nu\rho\sigma}=h^{\mu\rho}h^{\nu\sigma}+
h^{\mu\sigma}h^{\nu\rho}-\frac{2}{3}h^{\mu\nu}h^{\rho\sigma} \ .
\label{tau} \end{equation}

From the expression for the flux we see that, since $S$ is proportional
to $T^4$, there is the usual conduction term proportional to the
(four) gradient of $T$ with a conductivity coefficient $16\pi \bar
\varepsilon S/(3T)$.  This is the diffusion limit and it is much
less good than the Eddington approximation, for which the flux is
proportional to the gradient of $E$.   With the Eddington approximation,
thermal times are well approximated over all ranges of optical thickness
\citep{unn66}.  The second term in (\ref{flu}) has been considered to be
a purely relativistic term \citep{mih83}, but a classical analogue
exists in the kinetic theory of rarefied gases \citep{che00,bgk00}.

In the expression for the pressure tensor, the first term, to the order
given, does mimic the Eddington approximation and we have in addition
a viscous term.  From the standard form of the pressure tensor, we can
read off the viscosity coefficient given by Thomas, $16\epsilon \pi S/15$.
Like the Thomas expression for the flux, this is also a diffusive
approximation good only for short mean free paths.  Our aim here is to
improve on the non-diagonal viscous tensor in the same way that the
Eddington approximation improves on the conduction term.  In a subsequent
paper, we shall describe an improved approximation for the diagonal terms
for the case where scattering is important.

\section{The Expansion of $f$}\label{f}

To go beyond the Thomas approximation, we could continue the iteration
process described in the previous section, but we find it more effective
to proceed by formal expansions.  Our small parameter will be a typical
value of $\varepsilon$, call it $\epsilon$.  Then we form
the expansion
\begin{equation}
f=\sum_{m=0}^{\infty}f_{(m)}\epsilon^m \ ,
\label{exp}
\end{equation}
which we introduce into the transfer equation.  On demanding
that the expanded equation is satisfied term by term and
prescribing that $\varepsilon$ is everywhere of order $\epsilon$,
we get \begin{equation}
f_{(0)} ={\cal S} \, ,
\label{f0}
\end{equation}
as in comoving local thermodynamic equilibrium and, for the higher
orders, we obtain
\begin{equation}
f_{(m+1)}=-{\varepsilon \over \epsilon} n^{\mu}
f_{(m),\mu} \ . \label{fn}
\end{equation}

In the first order, as in diffusion theory, the complication caused by
having $\varepsilon$ in the formulas is slight, but in higher orders, it
makes the development very cumbersome especially on account of the
presence of derivatives of $\varepsilon$ in the expressions of $f_{(m)}$.
These may be converted to derivatives with respect to rest frequency and
state variables as for the source function but still, in the
performance of the integrals over frequency, it soon becomes necessary to
introduce a plethora of mean absorption coefficients, and even tensorial
mean absorption coefficients \citep{and72}. Though we do not report on
this here, we have retained variation of $\varepsilon$ in calcualtions
out to the third order in $\varepsilon$ in \citep{che00}.  This is far
enough for one aspect of our approach to be carried out, though we shall
not go into this here.

More generally, it is possible to make allowances for the dependence of
$\varepsilon$ on $x^\mu$ by a formal device like used in stellar
atmosphere theory.  We may introduce new coordinates $y^\mu$ through the
transformation $\epsilon dy^\mu = \bar \varepsilon d x^\mu$ where $\bar
\varepsilon$ is a mean (over frequency) absorption coefficient.  This
formal simplification is analogous to a transformation to the local rest
frame of the matter.  But since both $\bar \varepsilon$ and the $u^\mu$
are dependent on the radiation field, such implicit definitions of
coordinates necessitate, at the end of the calculations, some rather hard
inversions.  Moreover, we would need to take frequency means to
use such transformations effectively. We shall avoid all these
complications here and will stick to inertial coordinates, requiring
that $\varepsilon$ is constant. In that case, we obtain
\begin{equation}
f_{(m+1)} = -n^{\mu}f_{(m),\mu}
\, ,
\label{(fm)}
\end{equation}
or, in terms of ${\cal S}$,
\begin{equation}
f_{(m)} = (-1)^m{D^m {\cal S}\over Ds^m}
\, ,
\label{(m)}
\end{equation}
where \begin{equation}
{D\over Ds} = n^{\mu}\partial_{\mu} \, . \label{dds} \end{equation}

To perform the indicated differentiations, it is useful to have a
formula for the derivative of the null vector, $n^\mu$.  On
differentiating
$p^{\mu}=n^{\mu} \tilde\nu$ and making use of (\ref{commanu}), we
find that \begin{equation}
n^\mu_{\, ,\nu}=-n^\mu n^\rho u_{\rho,\nu} \ . \label{ncomma}
\end{equation}
We can then develop explicit formulas for the $f_{(m)}$ in terms of the
source function and these together lead to a series providing $f$ as a
functional of ${\cal S}$.  This is tantamount to what \citet{cha60} has
called the formal solution of the transfer equation. (Its
integral expression in terms of optical distance is helpful in treating
the case of variable $\varepsilon$.)  The idea of the formal solution is
that the source function and the absorption coefficient are material
properties of the medium and are to be considered as given for the
purposes of solving the transfer problem, though the situation is
reversed when we try to solve equations for the properties of the medium.
So we need to decide how the source function is to be specified.

In the simplest case mathematically, we would be given ${\cal S}$
explicitly as a function of space and time.  From this, we would compute
frequency integrals to get the integrated intensity through formula
(\ref{IntI}).  But it is more representative of the general situation
to assume, as we did for diffusion theory, that ${\cal S}$ is expressible
in terms of $T$ and $\tilde \nu$.  Though the source function does not
depend on direction, we do have effects of angular distribution entering
in through the transfer equation itself.  To make allowances for that
also, we write \begin{equation}
\partial_\mu = T_{,\mu}\,\partial_T+p^\rho
u_{\rho,\mu}\,\partial_{\tilde\nu}
+n^{\sigma}_{,\mu} \partial_{n^\sigma}
\label{partialmu}
\end{equation}
and we are led to a generalization of the differentiation formula
implicit in (\ref{intint1}):
\begin{equation}
\partial_\mu = \left(T_{,\mu}\, -  T n^\rho u_{\rho,\mu}\right)
\partial_T \ -n^\sigma n^\rho u_{\rho,\mu}\,\partial_{n^\sigma} \; .
\label{partialt2} \end{equation}

On noting that the quantity $f_{(0)} = {\cal S}$ does not depend on
direction ($n^\mu$) we can evaluate $f_{(1)}$.  We find using
(\ref{partialt2}) that
\begin{equation}
f_{(1)} = - {\cal S}_{,T} T_{,\mu} n^{\mu} + T {\cal S}_{,T}
 u_{\nu,\mu} n^{\mu} n^{\nu} \ .  \label{f1} \end{equation}
If we use this formula to compute the moments, we recover
the results of the previous section with $\epsilon$ in place
of $\bar \varepsilon$.

From (\ref{(m)}), we see on using (\ref{ncomma}), that \begin{equation}
f_{(2)}  = n^{\mu}n^{\nu}{\cal S}_{,\mu\nu}-n^{\mu}n^{\nu}n^{\rho}
u_{\rho,\mu}{\cal S}_{,\nu} \ . \label{IIf2} \end{equation}
We then have to cope with the higher derivatives of the
source function ${\cal S}$.  If we continue to assume that ${\cal S}$
depends only on $\tilde\nu/T$, since ${\cal S}$ does not depend on
$n^{\mu}$, (\ref{partialt2}) gives us \begin{equation}
{\cal S}_{,\mu}= T{\cal S}_{,T}
[(ln\, T)_{,\mu}-(ln\, \tilde\nu)_{,\mu}] \ .
\label{smu}
\end{equation}
A similar, but lengthier, calculation leads to \begin{eqnarray}
{\cal S}_{,\mu\nu} & = & \quad  T{\cal S}_{,T}\{[(ln T)_{,\mu}-
(ln\tilde\nu)_{,\mu}](ln T)_{,\nu}+
(ln T)_{,\mu\nu}-(ln\tilde\nu)_{,\mu\nu}\} \\
& + & [T^2{\cal S}_{,TT}(ln T)_{,\nu}-
(T{\cal S}_{,T}+T^2{\cal S}_{,TT})
(ln\tilde\nu)_{,\nu}]\left[(ln T)_{,\mu}-(ln\tilde\nu)_{,\mu}\right]
\nonumber \label{smunu}
\end{eqnarray}
where we have used (\ref{commanu}) and (\ref{ncomma}).

We then find that (\ref{IIf2}) takes the form \begin{equation}
f_{(2)}=n^{\mu}n^{\nu}{\cal F}^{(22)}_{\mu\nu}
-n^{\mu}n^{\nu}n^{\rho}{\cal F}^{(23)}_{\mu\nu\rho}
+n^{\mu}n^{\nu}n^{\rho}n^{\sigma}{\cal F}^{(24)}_{\mu\nu\rho\sigma}
\label{IIIf2} \end{equation}
with
\begin{eqnarray}
{\cal F}^{(22)}_{\mu\nu} & = & {\cal S}_{,T}T_{,\mu\nu}+
{\cal S}_{,TT}T_{,\mu}T_{,\nu}
\nonumber \\
{\cal F}^{(23)}_{\mu\nu\rho} & = & T{\cal S}_{,T}u_{\mu,\nu\rho}+(3{\cal
S}_{,T}+ 2T{\cal S}_{,TT})u_{\mu,\nu}T_{,\rho} \nonumber \\
{\cal F}^{(24)}_{\mu\nu\rho\sigma} & = & (3T{\cal S}_{,T}+
T^2{\cal S}_{,TT})u_{\mu,\nu}u_{\rho,\sigma}\ .
\label{IVf2} \end{eqnarray}

This way of writing $f_{(2)}$ is useful in computing terms in the
next order and from (\ref{(m)}) and (\ref{dds}) and the forms of
$f_{(1)}$ and $f_{(2)}$, we may surmise that \begin{equation}
f_{(m)} = \sum_{k=m}^{2m} (-1)^k {\cal F}^{(mk)}_{\mu_1
\cdots \mu_k} \,
n^{\mu_1}\cdots n^{\mu_k} \label{If3} \end{equation}
where the quantities ${\cal F}$ are functions of ${\cal S}$  and its
derivatives with respect to $T$; that is, they are functionals of
${\cal S}$.  (If we had not assumed constant photon mean free path,
they would be functionals of $\log \varepsilon$ as well.)  The calculation
of the distribution function at any order then boils down to the
calculation of these functionals, which is a straightforward, if
demanding, task.  For $m=3$, we are led to the results
\begin{equation} {\cal F}^{(33)}_{\mu\nu\alpha} = {\cal S}_{,TTT}T_{,\mu}
T_{,\nu}T_{,\alpha}+3{\cal S}_{,TT}
T_{,\mu\nu}T_{,\alpha}+{\cal S}_{,T}T_{,\mu\alpha\nu}
\end{equation}
\begin{eqnarray}
{\cal F}^{(34)}_{\mu\nu\alpha\beta} & = & T{\cal S}_{,T}
u_{\alpha,\beta\mu\nu}+(4{\cal S}_{,T}+
3T{\cal S}_{,TT})u_{\alpha,\beta\mu}T_{,\nu}+(6{\cal S}_{,T}+
3T{\cal S}_{,TT})u_{\alpha,\beta}T_{,\mu\nu} \nonumber \\
&   & + \; (8{\cal S}_{,TT}+3T{\cal
S}_{,TTT})u_{\alpha,\beta}T_{,\mu}T_{,\nu}
\end{eqnarray}
\begin{eqnarray}
{\cal F}^{(35)}_{\mu\nu\alpha\beta\rho}
& = & u_{\alpha,\beta}[(10T{\cal S}_{,T}+
3T^2{\cal S}_{,TT})u_{\mu,\nu\rho} \nonumber \\
&      & \ \ \ +(15{\cal S}_{,T}+18T{\cal S}_{,TT}+
3T^2{\cal S}_{,TTT})u_{\mu,\nu}T_{,\rho}]
\end{eqnarray}
\begin{equation}
{\cal F}^{(36)}_{\mu\nu\rho\sigma\alpha\beta}=[15T{\cal S}_{,T}+9T^2
{\cal S}_{,TT}+ T^3{\cal S}_{,TTT}]u_{\mu,\nu}
u_{\rho,\sigma}u_{\alpha,\beta}
\end{equation}
We thus have all the terms up to third order.

In this way we see that we may order the terms by increasing angular
complexity.  In that case, if we go up to four factors of $n^\mu$, we
have an expansion of $f$ in the form \begin{eqnarray}
f & = & {\cal S} - \epsilon {\cal F}^{(11)}_{\mu}n^\mu
  +\epsilon [{\cal F}^{(12)}_{\mu\nu}  + \epsilon
{\cal F}^{(22)}_{\mu\nu}] n^{\mu} n^{\nu}
-\epsilon^2 [{\cal F}^{(23)}_{\mu\nu\rho} +
\epsilon {\cal F}^{(33)}_{\mu\nu\rho}]
               n^{\mu}n^{\nu}n^{\rho} \nonumber \\
  &   & -\epsilon^2 [{\cal F}^{(24)}_{\mu\nu\rho\sigma} +
                    \epsilon {\cal F}^{(34)}_{\mu\nu\rho\sigma}
       + \epsilon^2 {\cal F}^{(44)}_{\mu\nu\rho\sigma}]
                n^{\mu}n^{\nu}n^{\rho}n^{\sigma}
+ \cdots.  \label{fm} \end{eqnarray}
We may compute the coefficients ${\cal F}$ at any order using this
procedure.

\section{The Intensity Expansion}

An intermediate step on the way to computing the moments that make up the
radiative stress tensor, is the computation of the (frequency) integrated
intensity defined in (\ref{IntI}).  Its expansion is \begin{equation}
I = \sum_{m=0}^\infty \epsilon^m I_{(m)} \, ,\label{Iexp} \end{equation}
and when we introduce (\ref{If3}) into (\ref{IntI}) we obtain for the
coefficients in (\ref{Iexp}),  \begin{equation}
I_{(m)}=\sum_{k=m}^{2m}(-1)^k Y_{\mu_1\cdots \mu_k}^{(m,k)} \,
n^{\mu_1}\cdots n^{\mu_k}     \label{wise}
\end{equation}
where
\begin{equation}
Y_{\mu_1\cdots \mu_k}^{(m,k)}=\int\tilde{\nu}^3
{\cal F}_{\mu_1\cdots \mu_k}^{(mk)} d\tilde{\nu} \label{Y} \; .
\end{equation}

With the expressions for the ${\cal F}_{\mu_1\cdots \mu_k}^{(mk)}$
given in the previous section, we may develop explicit expressions for
$Y_{\mu_1\cdots \mu_k}^{(m,k)}$.  We do this by working out a recursion
formula for these coefficients.  First, we introduce the expression
(\ref{If3}) for $f_{(m)}$ into (\ref{(fm)}) to obtain \begin{equation}
f_{(m+1)}  =  - n^\mu \sum_{k=m}^{2m} (-1)^{k} \left[
{\cal F}^{(mk)}_{\mu_1 \cdots \mu_k \, , \mu}  n^{\mu_1} \cdots
n^{\mu_{k}} +  {\cal F}^{(mk)}_{\mu_1 \cdots \mu_k}
(n^{\mu_1} \cdots n^{\mu_{k}})_{,\mu} \right] \; .
\label{diffF} \end{equation}
Since from (\ref{ncomma}), we have \begin{equation}
(n^{\mu_1} \cdots n^{\mu_{k}})_{,\mu} = - k\, n^\rho\,
u_{\rho,\mu} (n^{\mu_1} \cdots n^{\mu_{k}})\; ,  \end{equation}
the only real complication comes from the derivatives of the ${\cal F}$.

Next, we reexamine the derivation of the explicit formulae for the
${\cal F}$, and see that their dependence on position and time comes
about through their dependence on $\tilde \nu$, on $T$ and on
$u^\mu$, as well as the derivatives of these quantities with respect
to $x^\mu$.  So each ${\cal F}$ depends on $\tilde \nu$ and on $N$ other
variables that we shall call $V^A$, where $A=1,2,\dots,N$.
The $V^A$ are $T_{,\mu}$, $T_{,\mu\nu}$, ..., $u_{\mu,\nu}$,
and so on.  Let \begin{equation}
\Lambda^{(m,k)}_{\; \mu} = V^A_{\; ,\mu}\; \partial_{V^A}
\end{equation}
where summation over repeated $A$ is understood.
Then
\begin{equation}
\partial_\mu = \tilde \nu_{,\mu} \partial_{\tilde \nu} +
\Lambda^{(m,k)}_{\;\mu} \label{nonu} \end{equation}
On recalling that $\tilde \nu_{,\mu} = \tilde \nu n^\rho u_{\rho,\mu}$
we can rewrite (\ref{diffF}) as
\begin{eqnarray}
f_{(m+1)} & = & \sum_{k=m}^{2m} (-1)^{k+1}
\tilde \nu n^\rho n^\mu u_{\rho,\mu} \partial_{\tilde \nu}
{\cal F}^{(mk)}_{\mu_1 \cdots \mu_k} n^{\mu_1} \cdots n^{\mu_{k}}
\nonumber \\
& + & \sum_{k=m}^{2m} (-1)^{k+1} \left[n^\mu \Lambda^{(mk)}_{\; \mu}
- k n^\rho n^\mu u_{\rho,\mu}\right]
{\cal F}^{(mk)}_{\mu_1 \cdots \mu_k} n^{\mu_1} \cdots n^{\mu_{k}}
\end{eqnarray}

To compute the $I_{(m)}$, we carry out the frequency integral
indicated in (\ref{IntI}) and integrate by parts in the term with
$\partial_{\tilde \nu}$ to obtain
\begin{equation}
I_{(m+1)}  =  \sum_{k=m}^{2m} (-1)^{k}\left[
(k+4) n^\rho n^\mu u_{\rho,\mu}  -  n^\mu \Lambda^{(m,k)}_{\; \mu}\right]
Y^{(m,k)}_{\mu_1 \cdots \mu_k} n^{\mu_{1}} \cdots n^{\mu_{k}} \; .
\label{2term}  \end{equation}
Now $\tilde \nu$ is no longer in the problem, and so (\ref{nonu})
simplifies to $\Lambda^{(m,k)}_{\; \mu} = \partial_\mu$; we shall make
this replacement in the rest of the development.

Formula (\ref{2term}) has two terms, each having a different
angular structure.  To keep to our general attempt to order things by
angular structure as well as by mean free path, we make a
further rearrangement.  So far, in the expressions for the ${\cal F}$,
for different pairs $(mk)$, we have had need only of values
for which $m \le k \le 2m$.  Therefore we are at liberty to impose that
\begin{equation}
{\cal F}^{(mk)}_{\mu_1 \cdots \mu_k} = 0 \qquad
{\rm for} \qquad k<m \qquad {\rm and} \qquad k>2m\; . \label{cond}
\end{equation} Accordingly  \begin{equation}
Y^{(m,k)}_{\mu_1 \cdots \mu_k} = 0 \qquad
{\rm for} \qquad k<m \qquad {\rm and} \qquad k>2m\; . \label{condy}
\end{equation}  Moreover, for either $m<0$ or $k<0$, we
define $Y^{(m,k)}_{\mu_1 \cdots \mu_k}$ to be zero.

Now we can adjust the indices in the summation (\ref{2term}) by
replacing $k$ in the first term with $k-2$ and $k$ in the second term with
$k-1$.  We then have
\begin{equation}
I_{(m+1)}  =  \sum_{k=m+1}^{2(m+1)} (-1)^{k} \left[
(k+2) Y^{(m,k-2)}_{\mu_1 \cdots \mu_{k-2}}  \;
u_{\mu_{k-1},\mu_k} +  Y^{(m,k-1)}_{\mu_1 \cdots \mu_{k-1}\, , \mu_k}
\right] n^{\mu_1} \cdots n^{\mu_k}
\end{equation}
and, on comparing with (\ref{wise}), we observe that
\begin{equation}
\label{Yrecursion}
Y^{(m+1,k)}_{\mu_1 \cdots \mu_k}  =
Y^{(m,k-1)}_{\mu_1 \cdots \mu_{k-1},\mu_k}
+ (k+2) Y^{(m,k-2)}_{\mu_1 \cdots \mu_{k-2}}\; u_{\mu_{k-1},\mu_k} \; ,
\label{recurs} \end{equation}
which will be used in the sequel.   The $Y$ sequence is to be built up
from $Y^{(0,0)}$.  As we see from (\ref{Y}), this is the integrated
source function, that is \begin{equation}
Y^{(0,0)} = S.  \label{00} \end{equation}
Hence, as seen above, $I$ is a functional of $S$ as well as of
$u^\mu$ and the angular variables.

\section{The Viscosity Tensor}

Once we have calculated the integrated intensity, as in the
previous section, we can evaluate the physically important angular
moments, \begin{equation}
E=\sum_{m=0}^\infty \epsilon^m E_{(m)}\, , \quad
F^\mu=\sum_{m=0}^\infty \epsilon^m F_{(m)}^\mu \, , \quad
P^{\mu\nu}=\sum_{m=0}^\infty \epsilon^m P^{\mu\nu}_{(m)}\, .
\label{expm} \end{equation}
Since the $Y$ defined in (\ref{Y}) are independent of angle, the
terms in these expansions are  \begin{equation}
E_{(m)} = \sum_{k=m}^{2m}(-1)^k Y_{\mu_1\cdots \mu_k}^{(m,k)}
\int n^{\mu_1} n^{\mu_2} \cdots n^{\mu_k}\; d\Omega \label{E(m)}
\end{equation} \begin{equation}
F_{(m)}^\mu = \sum_{k=m}^{2m}(-1)^k Y_{\mu_1\cdots \mu_k}^{(m,k)}
\int l^\mu n^{\mu_1} n^{\mu_2} \cdots n^{\mu_k}\; d\Omega \label{F(m)}
\end{equation} \begin{equation}
P^{\mu\nu}_{(m)} = \sum_{k=m}^{2m}(-1)^k Y_{\mu_1\cdots \mu_k}^{(m,k)}
\int l^\mu l^\nu n^{\mu_1} n^{\mu_2} \cdots n^{\mu_k}\; d\Omega \; .
\label{P(m)} \end{equation}

Though we shall quote some results about the approximations to $E$
and $F^\mu$, our main interest is in providing an improved treatment of
the viscous effects for transparent regions of the system under study.
So we shall detail only the calculation of an improved approximation
for the pressure tensor.  If we introduce the Eddington closure for
the diagonal terms in the stress tensor, we may write
\begin{equation}
P^{\mu\nu} = - {1\over 3} E h^{\mu\nu} + \Xi^{\mu\nu} \; ,
\label{visc} \end{equation}
where $\Xi^{\mu\nu}$ is the viscous stress tensor, whose expression
the main goal of this work.  As mentioned, in a later paper, we
shall show how the first term in (\ref{visc}) can be improved
considerably.

To seek a representation for $\Xi^{\mu\nu}$ that is suitable for
both long and short photon mean free paths, we begin by expanding the
viscosity tensor as \begin{equation}
\Xi^{\mu\nu} = \sum_{m=0}^\infty \epsilon^m \Xi^{\mu\nu}_{(m)} \; .
\label{xi2}
\end{equation}
From (\ref{E(m)}) and (\ref{P(m)}) we find the coefficients
in this expansion and so can write \begin{equation}
\Xi^{\mu\nu}_{(m)} =  \sum_{k=m}^{2m}(-1)^k Y_{\mu_1\cdots
\mu_k}^{(m,k)} {\mit\Omega}^{\mu\nu\mu_1\cdots\mu_{k}},
\label{Xi(m)}
\end{equation}
where
\begin{equation}
{\mit\Omega}^{\mu\nu\mu_1\cdots\mu_{k}} \equiv \int (l^\mu l^\nu +
 {1\over 3}h^{\mu\nu})
n^{\mu_1} n^{\mu_2} \cdots n^{\mu_k}\; d\Omega \; . \label{Omega}
\end{equation}
Since the trace of $l^\mu l^\nu + {1\over 3}h^{\mu\nu}$ is zero,
$\Xi^{\mu\nu}_{(m)}$ and $\Xi^{\mu\nu}$, are traceless, symmetric
tensors.

For the $Y$s we have the recursion formula (\ref{recurs})
given at the end of the previous section,
while to reexpress the angular integrals usefully,
we can refer to the formula of Appendix A.  We see from these that
\begin{equation} \Omega^{\mu\nu}=0, \qquad\quad \Omega^{\mu\nu\rho} = 0
\qquad\quad {\rm and} \qquad\quad
\Omega^{\mu \nu \rho \sigma} = \frac{4\pi}{15} \tau^{\mu \nu \rho \sigma}
\ , \label{Om=0} \end{equation}
where $\tau^{\mu\nu\rho\sigma}$ is defined in (\ref{tau}).
To extend the results of Thomas we next generalize
$\tau^{\mu\nu\rho\sigma}$ into a series of orthogonal
polynomials that represent the expansion of $\Omega$.

If we use (\ref{bino}), we can develop the products of the $n^\mu$ and
express $\Omega^{\mu\nu\mu_1\cdots \mu_{k}}$ in terms of the moments
of $l^\mu$.  Then we have \begin{equation}
{\mit\Omega}^{\mu\nu\mu_1 \cdots \mu_{k}} =
\sum_{i=0}^{k} \pmatrix{i\cr k\cr}
W^{\mu\nu(\mu_{1}\cdots\mu_{i}}u^{\mu_{i+1}}
\cdots u^{\mu_{k})} \label{Om0} \end{equation}
where \begin{equation}
W^{\mu\nu\mu_1\dots\mu_\ell} =
M^{\mu\nu\mu_1\dots\mu_\ell} + {1\over 3}h^{\mu\nu}
M^{\mu_1\dots\mu_\ell}    \end{equation} and  \begin{equation}
M^{\mu_1\dots\mu_\ell} = \int l^{\mu_1} l^{\mu_2}\cdots
l^{\mu_\ell} d\Omega \ . \label{ang} \end{equation}

The integral for an odd number of factors $l^\mu$ is zero so,
for odd $k$, $M^{\mu_1\dots\mu_k}\equiv 0$, and hence
$W^{\mu\nu\mu_1\dots\mu_k}\equiv 0$.  Therefore
we may rewrite (\ref{Om0}) in a way that will be convenient
later, namely, \begin{equation}
{\mit\Omega}^{\mu\nu\mu_1 \cdots \mu_{k}} =
\sum_{i=0}^{[\frac{k}{2}]} \pmatrix{2i\cr k\cr}
W^{\mu\nu(\mu_{1}\cdots\mu_{2i}}u^{\mu_{2i+1}} \cdots u^{\mu_k)}
 \label{Om} \end{equation}
where $[\frac{k}{2}]$ means the integer part of $\frac{k}{2}$.

When $k=1$, the upper limit of the sum is $i=0$,
and $\Omega^{\mu\nu\rho} = W^{\mu\nu}u^{\rho}$. Because
$M=4\pi$ and, according to (\ref{M2}),
$M^{\mu\nu}=-{4\pi\over 3}h^{\mu\nu}$, we have $W^{\mu\nu}=0$
and $\Omega^{\mu\nu\rho} = 0$ as just noted.  In the same way,
at the next level, from (\ref{even}) we see that \begin{equation}
\tau^{\mu \nu \rho \sigma} = W^{\mu \nu \rho \sigma} \label{newt}
\end{equation}
which is equivalent to the third expression in (\ref{Om=0}).
We then define a generalized $\tau$ as \begin{equation}
\tau^{\mu\nu\mu_{1}\cdots\mu_{2\ell}} =
W^{\mu\nu\mu_1\dots\mu_{2\ell}}
+ {1\over 3} W^{\mu\nu\mu_1\dots\mu_{2\ell-3}\mu_{2\ell-2}}
h^{\mu_{2\ell-1}\mu_{2\ell}} \ .     \label{gtau} \end{equation}
To introduce this into (\ref{Om}) we rewrite (\ref{gtau}) as
\begin{equation}
W^{\mu\nu\mu_1\dots\mu_{2\ell}} =
\tau^{\mu\nu\mu_{1}\cdots\mu_{2\ell}} -
{1\over 3} W^{\mu\nu\mu_1\dots\mu_{2\ell-3}\mu_{2\ell-2}}
h^{\mu_{2\ell-1}\mu_{2\ell}} \ ,
 \label{gW} \end{equation}
where it is understood that $h^{\mu_p \mu_q}$ is zero if either
$p$ or $q$ is negative.  Hence, for $\ell=1$ we recover (\ref{newt})
and we may then proceed recursively to express $W$ in terms
of $\tau$.  When $\ell = 2$ we have \begin{equation}
W^{\mu \nu \mu_1 \mu_2 \mu_3 \mu_4 }=
\tau^{\mu \nu \mu_1 \mu_2 \mu_3 \mu_4} - {1\over 3}
\tau^{\mu \nu \mu_1 \mu_2} h^{\mu_3 \mu_4} \ . \end{equation}
From this we see that \begin{equation}
\label{Wtau}
W^{\mu\nu\mu_1\cdots\mu_{2p}} = \sum_{q=0}^{p}
\left(-{1\over 3}\right)^{p-q}\tau^{\mu\nu\mu_1\cdots\mu_{2q}}
h^{\mu_{2q+1}\mu_{2q+2}}\cdots h^{\mu_{2p-1}\mu_{2p}}\; . \label{wha}
\end{equation}
When we put (\ref{wha}) into (\ref{Om}) we obtain the desired expansion
of $\Omega$ in terms of the generalized $\tau$: \begin{eqnarray}
\Omega^{\mu\nu\mu_1\cdots\mu_k} & = & \sum_{i=0}^{[\frac{k}{2}]}
\pmatrix{2i \cr k} \sum_{j=0}^{i}\left(-\frac{1}{3}\right)^{i-j}
\nonumber \\ &\times&
\tau^{\mu\nu(\mu_1\cdots\mu_{2j}}h^{\mu_{2j+1}}\cdots
h^{\mu_{2i-1}\mu_{2i}}
\, u^{\mu_{2i+1}}\cdots u^{\mu_k)} \ .\label{om1.0}
\end{eqnarray}
The generalized $\tau$ are useful for developing the angular
structure of tensors and we find them more suited to this problem than
the spherical harmonics of the usual moment schemes.

\section{Ordering the Summations}

In a calculation that is in some ways the predecessor of this one
\citep{unn66}, a series of angular moments was rearranged
and resummed to provide a closure approximation for the thermal terms.
These terms lacked off-diagonal elements and so gave only the Eddington
approximation, which is nevertheless a great improvement on the diffusion
approximation.  The moral is that it is very effective to ensure that all
orders in mean free path made themselves felt in the result.  However,
since the approximation used in that earlier work fails to capture the
viscous effects, we here propose one that retains the off-diagonal terms
and leads to results good both for long and short photon mean free paths.
To bring out the nature of the rearrangement we propose, let us shift the
index $k$ in (\ref{Xi(m)}) to $k+m$ so that (\ref{xi2})-(\ref{Xi(m)})
becomes \begin{equation}
\Xi^{\mu\nu} = \sum_{m=0}^{\infty} \sum_{k=0}^m
\Upsilon^{\mu\nu}_{(m,k)}  \label{upsum} \end{equation}
where \begin{equation}
\Upsilon^{\mu\nu}_{(m,k)}  = (-1)^{k+m} \epsilon^m
Y^{(m,m+k)}_{\mu_1 \cdots \mu_{k+m}} \Omega^{\mu\nu\mu_1 \cdots \mu_{k+m}}
\; . \label{ups} \end{equation}

When we write out the $\Upsilon^{\mu\nu}_{m,k}$ as an array of tensors,
with $m$ designating the rows and $k$ designating the columns, it takes
the form \begin{equation}
\begin{array}{cccc}
\Upsilon^{\mu\nu}_{(0,0)} & 0 & 0 & \cdots \\
\Upsilon^{\mu\nu}_{(1,0)} & \Upsilon^{\mu\nu}_{(1,1)} & 0 & \vdots \\
\Upsilon^{\mu\nu}_{(2,0)} & \Upsilon^{\mu\nu}_{(2,1)} &
\Upsilon^{\mu\nu}_{(2,2)} & \ddots \\
\cdots & \cdots & \cdots & \ddots
\end{array}
\end{equation}
Thus, the $\Upsilon^{\mu\nu}_{(m,k)}$ form a lower triangular
matrix of tensors.

The first sum in (\ref{upsum}) is taken over the $m^{\rm th}$ row in
the array of $\Upsilon$s and the second sum is over the columns.
In the spirit of the earlier derivation of the Eddington approximation by
resummation, we rearrange the terms so that the sum is first taken over
columns, then over the rows.  That is, we can sum first over the $m$.
Since, for fixed $k$, the sum over $m$ begins with the value $m=k$, it is
then convenient to shift the index $m$ by $k$.  This simple rearrangement
amounts to \begin{equation}
\sum_{m=0}^\infty \sum_{k=0}^m \Upsilon^{\mu\nu}_{(m,k)} =
\sum_{k=0}^\infty \sum_{m=0}^\infty \Upsilon^{\mu\nu}_{(m+k,k)}\; .
\label{trick} \end{equation}
This device, which we shall employ again, turns (\ref{upsum}) into
\begin{equation}
\Xi^{\mu\nu} = \sum_{k=0}^\infty \sum_{m=0}^\infty
\Upsilon^{\mu\nu}_{(m+k,k)} \label{cr} \end{equation} so that the
expansion of the viscosity tensor is now \begin{equation}
\Xi^{\mu\nu} = \sum_{k=0}^{\infty} \epsilon^k \Xi^{\mu\nu}_{[k]}
\label{Pi} \end{equation} with \begin{equation}
\Xi^{\mu\nu}_{[k]} = \sum_{m=0}^{\infty} (-1)^m \epsilon^m
Y_{\mu_1 \cdots \mu_{m+2k}}^{(m+k,m+2k)}
\Omega^{\mu\nu\mu_1 \cdots \mu_{m+2k}} \; . \label{Pisum} \end{equation}

The two expansions for $\Xi^{\mu\nu}$, (\ref{xi2}) and (\ref{Pi})
look similar but, when we compare them carefully, we detect an important
difference: only the series for $\Xi^{\mu\nu}_{[k]}$ contains terms of all
orders in $\epsilon$.  We see also that (\ref{cr}) involves
$\Upsilon^{\mu\nu}_{(m+k,k)}$, which in
turn requires that we be more explicit about the $Y_{\mu_1 \cdots
\mu_{m+2k}}^{(m+k,m+2k)}$ in (\ref{Pisum}).  The relevant development can
proceed with the help of the iteration formula (\ref{recurs}).  We
first use (\ref{condy}) in (\ref{recurs}) with $m \rightarrow m-1$ and
$k=m$ to obtain \begin{equation}
Y^{(m,m)}_{\mu_1 \cdots \mu_m} =
Y^{(m-1,m-1)}_{\mu_1 \cdots \mu_{m-1}, \mu_m} \ . \end{equation}
Iteration of this formula gives \begin{equation}
Y^{(m,m)}_{\mu_1 \cdots \mu_m} =
Y^{(0,0)}_{,\, \mu_1 \cdots \mu_m}.  \label{k=0} \end{equation}

More generally, for $k>0$,
we may use $(\ref{recurs})$ to find \begin{equation}
Y^{(k,2k)}_{\mu_1\cdots \mu_{2k}} =
(2k+2) Y^{(k-1,2k-2)}_{\mu_1\cdots \mu_{2k-2}} u_{\mu_{2k-1},\mu_{2k}}
\; . \label{737}  \end{equation}
Again, using (\ref{recurs}), we see that \begin{equation}
Y^{(k+1,2k+1)}_{\mu_1\cdots \mu_{2k+1}} =
Y^{(k,2k)}_{\mu_1\cdots \mu_{2k}, \mu_{2k+1}} +
(2k+3) Y^{(k,2k-1)}_{\mu_1 \cdots \mu_{2k-1}} u_{\mu_{2k}, \mu_{2k+1}}
\; , \label{738} \end{equation}
and so on.  For the special case of $k=0$, we obtain (\ref{k=0}).

Now we can use the rearrangement trick again by noting that if we put
(\ref{737})-(\ref{738}) into (\ref{Pisum}), it takes the form
\begin{equation}
\Xi^{\mu\nu}_{[k]} =
\sum_{m=0}^\infty \sum_{i=0}^m X^{\mu\nu}_{(m+k,k,i)}
\end{equation}
where the expressions for the $X$ can be obtained by comparing to
(\ref{Pisum}).  We do not need their explicit form since we need
only to observe that by a repetition of the formula (\ref{trick}) we
get \begin{equation}
\Xi^{\mu\nu}_{[k]} =
\sum_{i=0}^\infty \sum_{m=0}^\infty X^{\mu\nu}_{(m+k+i,k,i)}
\ .  \end{equation}
We may then carry out the necessary substitutions, and separate out
the exceptional case $k=0$, to obtain \begin{eqnarray}
\Xi^{\mu\nu}&=& \sum_{k=1}^\infty \sum_{i=0}^\infty \sum_{m=0}^\infty
(-1)^{m+i}\epsilon^{m+k+i} (2k+i+2)
\Omega^{\mu\nu\mu_1\cdots\mu_{2k+m+i}}  \nonumber \\
&  &\times\left[Y^{(k+i-1,2k+i-2)}_{\mu_1\cdots\mu_{2k+i-2}} \,
u_{\mu_{2k+i-1},\mu_{2k+i}}\right]_{,\mu_{2k+1+i}\cdots\mu_{2k+m+i}}
\nonumber \\
&  &+\sum_{m=0}^\infty(-1)^m\epsilon^m Y_{,\mu_1 \cdots \mu_m}^{(0,0)}
\Omega^{\mu\nu\mu_1\cdots\mu_m}  \label{YYY}
\end{eqnarray}

This completes the formal expansion procedure for the viscosity
tensor, which we can now express in terms of the generalized $\tau$
and the coefficients of the intensity expansion.  For both of these
sets of quantities we have recursion formulae.  We next begin to make
approximations to prepare the way for representing the results in terms
of rational approximations to the series.

\section{Approximating the Orders}

So far we have been proceeding with formal expansions in mean free
path and making convenient rearrangements of the series obtained.  The
result is that terms in the series depend on $\epsilon$ so that terms
of similar character can be compared side by side.  Now we turn to some
approximations to make the series more manageable.  What we shall propose
in this section is the neglect of those terms in each order of the formal
expansion that are clearly (and asymptotically) smaller than their
fellows in the same order, as in the earlier study of the radiative heat
equation.  There it was observed that by retaining a dominant term in
every order one is led to a useful outcome (in that case, the Eddington
approximation).  That approach is what we mean by `approximating the
orders.'

We note that in (\ref{YYY}) we may
change the order of summation, since the sums are each taken to infinity
in that expression.  So we may rewrite (\ref{YYY}) more compactly as
\begin{equation}
\Xi^{\mu\nu}= \sum_{m=0}^\infty(-\epsilon)^m\biggl \{
\sum_{k=1}^\infty \sum_{i=0}^\infty
(-1)^i \epsilon^{k+i} A^{\mu\nu}_{(m,k,i)}
+ Y_{,\mu_1 \cdots \mu_m}^{(0,0)}
\Omega^{\mu\nu\mu_1\cdots\mu_m}\biggr \} \label{YY}
\end{equation} where
$$A^{\mu\nu}_{(m,k,i)}=(2k+i+2)
\Omega^{\mu\nu\mu_1\cdots\mu_{2k+m+i}}
\left[Y^{(k+i-1,2k+i-2)}_{\mu_1\cdots\mu_{2k+i-2}}
u_{\mu_{2k+i-1},\mu_{2k+i}}\right]_{,\mu_{2k+1+i}\cdots\mu_{2k+m+i}}\ .
$$
\begin{equation}\label{defA}\end{equation}

We operate first on the double sum of the summand of (\ref{YY}) by
separating out its leading term for $k=1$ and $i=0$.  The double sum can
then be rewritten as \begin{equation}
\epsilon A^{\mu\nu}_{(m,1,0)} +\sum_{k=2}^\infty \sum_{i=1}^\infty(-1)^i
\epsilon^{k+i}A^{\mu\nu}_{(m,k,i)}+\sum_{k=2}^\infty
\epsilon^k A^{\mu\nu}_{(m,k,0)}+\sum_{i=1}^\infty (-1)^{i} \epsilon^{i+1}
A^{\mu\nu}_{(m,1,i)}
\ .
\label{1st1}
\end{equation}
In the double sum of (\ref{1st1}), we shift the indices so that $k$
becomes $k+2$ and $i$ goes over into $i+1$.  Similar shifts
can be carried out for the last two terms of sum.  Then (\ref{1st1})
becomes
\begin{equation}
\epsilon A^{\mu\nu}_{(m,1,0)}
-\epsilon^3\sum_{k=0}^\infty \sum_{i=0}^\infty
(-1)^i\epsilon^{k+i}A^{\mu\nu}_{(m,k+2,i+1)}+\epsilon^2
\sum_{k=0}^\infty \epsilon^k \left[A^{\mu\nu}_{(m,k+2,0)}-(-1)^{k}
A^{\mu\nu}_{(m,1,k+1)}\right]\ .  \label{1st2} \end{equation}
Since the last two terms in this expression are down by at least a factor of
$\epsilon$ from the first term, from which they are otherwise not
dissimilar, we may neglect them.

Next, let us look at the second term in the summand of (\ref{YY}).
A simplification of this term can be based on the fact, seen on
inspection, that $\Omega^{\mu\nu}=0$, together with the result given in
(\ref{Om=0}) that $\Omega^{\mu\nu\rho}=0$.  With the first two terms
gone from this second sum over $m$ it becomes \begin{equation}
\sum_{m=0}^\infty(-\epsilon)^m Y_{,\mu_1 \cdots \mu_m}^{(0,0)}
\Omega^{\mu\nu\mu_1\cdots\mu_m}=\sum_{m=2}^\infty(-\epsilon)^m Y_{,\mu_1
\cdots \mu_m}^{(0,0)}
\Omega^{\mu\nu\mu_1\cdots\mu_m}  \ .    \label{107}
\end{equation}
By shifting $m$ to $m+2$, we can turn (\ref{107}) into
\begin{equation}
\sum_{m=0}^\infty(-\epsilon)^{m+2} Y_{,\mu_1 \cdots \mu_{m+2}}^{(0,0)}
\Omega^{\mu\nu\mu_1\cdots\mu_{m+2}} \ ,
\end{equation}
and so, we can write (\ref{YY}) as
\begin{eqnarray}
\Xi^{\mu\nu}  =  \sum_{m=0}^\infty(-\epsilon)^m
\biggl \{\epsilon A^{\mu\nu}_{(m,1,0)}
&-& \epsilon^3\sum_{k=0}^\infty \sum_{i=0}^\infty
(-1)^i\epsilon^{k+i}A^{\mu\nu}_{(m,k+2,i+1)}  \nonumber \\
&+&\left. \epsilon^2 Y_{,\mu_1\cdots \mu_{m+2}}^{(0,0)}
\Omega^{\mu \nu \mu_1 \cdots \mu_{m+2}}\biggr\}  \right. \ .
\label{xiapp1} \end{eqnarray}
As we have stated, we are going to neglect the middle
term in this summand and now we see that we may also neglect
the last term.  Our expression for the radiative viscous
stress tensor may then be approximated as \begin{equation}
\Xi^{\mu\nu} = 4\epsilon \sum_{m=0}^\infty(-\epsilon)^m
\Omega^{\mu\nu\rho\sigma\mu_1\cdots\mu_m}[Y^{(0,0)}u_{\rho,\sigma}]
_{,\mu_1\cdots\mu_m}
\label{xiapp}
\end{equation}
where each of the neglected terms is $O(\epsilon)$ compared to its
counterparts.  We note that the term for $m=0$ of this expansion is
the Thomas approximation to the stress tensor.

\section{The Viscous Smoothing Operator \label{vso}}

The thermal smoothing operator in (\ref{diff}) is a rational fraction of
polynomials in the Laplacian operator.  It was obtained by first
expanding the original integral operator of radiative smoothing in a
series of powers of photon mean free path and approximating each term
according to various criteria, especially simplicity of geometrical
structure.  In attempting a similar approach here for the viscosity
problem we encounter a more complicated development.  We have grouped
terms in our expansion by a combination of order in the mean free path
expansion and geometric complexity.  This approach has permitted us to
compare terms of like character and then to keep in each cluster of terms
the leading ones in powers of mean free path.  Now we wish to examine the
remaining terms and keep only the geometrically simplest terms, as was
done in the derivation of ${\cal D}$ of (\ref{diff}).  We shall also make
some further approximations in terms of $\epsilon$.

As always, much depends on $Y^{(0,0)} = S$  (see (\ref{00})), which we
now introduce into (\ref{xiapp}).  We then observe that the
viscous stress tensor takes the simple form
\begin{equation}
\Xi^{\mu\nu} = 4\epsilon \DD^{\mu\nu\rho\sigma}
\left[Su_{\rho,\sigma}\right] \end{equation}
where the smoothing operator, $\DD$, the centerpiece of this
investigation, is given as \begin{equation}
\DD^{\mu\nu\rho\sigma} = \sum_{m=0}^\infty
(-\epsilon)^m \Omega^{\mu\nu\rho\sigma\mu_1\cdots \mu_{m}}
\partial_{\mu_{1}\cdots \mu_{m}} \ .  \label{DD} \end{equation}
We now wish to extract the geometrically simplest part from each
term in this expansion.   To do this, we recall formula (\ref{om1.0})
for $\Omega^{\mu\nu\mu_1\cdots\mu_m}$, in which the leading
nonvanishing term in the series involves the tensor
$\tau^{\mu\nu\rho\sigma}$ that also appears in the standard form
of the viscous stress tensor.   The origin of the general $\tau$ has been
discussed in the derivation of (\ref{Wtau}).

To obtain the usual form of the Eddington approximation, one normally
omits higher order angular moments of the intensity field in
series such as we have obtained.  But the conventionally neglected
moments contain projections onto $\tau^{\mu\nu\rho\sigma}$ and so that
form of truncation misrepresents the viscosity tensor.  That is why we
developed the $\tau$ tensors --- by retaining only the leading angular
approximation in these quantities we keep just the angular quantity that
is central to the representation of viscosity.  Therefore, we here select
from (\ref{om1.0}) only the geometrically most elementary part of the term
in each order and write \begin{equation}\label{om2}
\Omega^{\mu\nu\mu_1\cdots\mu_m}=
\sum_{\ell=1}^{[\frac{m}{2}]}\pmatrix{2\ell \cr m}
\left(-\frac{1}{3}\right)^{\ell-1}
\tau^{\mu\nu(\mu_1\mu_2}
h^{\mu_3\mu_4}\cdots h^{\mu_{2\ell-1}\mu_{2\ell}}
\cdot u^{\mu_{2\ell-1}}\cdots u^{\mu_m)} \ .
\end{equation}
In allowing contributions to each term from only the leading generalized
$\tau$ we have made what may be called a Galerkin truncation.

Next we make an index shift $j \rightarrow \ell-1$, so that (\ref{om2})
becomes \begin{equation}
\Omega^{\mu\nu\mu_1\cdots\mu_m}  =
\sum_{j=0}^{[\frac{m-2}{2}]}\pmatrix{2j+2 \cr m}
\left(-\frac{1}{3}\right)^j
\tau^{\mu\nu(\mu_1\mu_2}h^{\mu_3\mu_4}\cdots
h^{\mu_{2j+1}\mu_{2j+2}}
u^{\mu_{2j+3}}\cdots u^{\mu_m)} \; .
\end{equation}
This in turn can be written as \begin{equation}
\Omega^{\mu\nu\rho\sigma\mu_1\cdots\mu_m}=
\sum_{j=0}^{[\frac{m}{2}]}
\pmatrix{2j+2 \cr m+2}\left(-\frac{1}{3}\right)^j
\tau^{\mu\nu(\rho\sigma}h^{\mu_1\mu_2}\cdots h^{\mu_{2j-1}\mu_{2j}}
u^{\mu_{2j+1}}\cdots u^{\mu_m)} \; .
\label{om4} \end{equation}
With this approximation for $\Omega$, the viscous smoothing operator is
\begin{eqnarray}
\DD^{\mu\nu\rho\sigma} & = &  \sum_{m=0}^\infty
(-\epsilon)^m
\sum_{j=0}^{[\frac{m}{2}]}\pmatrix{2j+2
\cr m+2}\left(-\frac{1}{3}\right)^j
\nonumber \\ &\times &
\tau^{\mu\nu(\rho\sigma}
h^{\mu_1\mu_2}\cdots h^{\mu_{2j-1}\mu_{2j}}
u^{\mu_{2j+1}}\cdots u^{\mu_m)}\partial_{\mu_{1}
\cdots \mu_{m}} \; . \label{DD1}\end{eqnarray}

Because $\partial_{\mu_{1}\cdots \mu_{m}}$ is symmetric,
the symmetrization of the $\mu_i$ is redundant and may be omitted.
However, this is not the case for indices $\rho$ and $\sigma$. If
we approximate (\ref{DD1}) by keeping only the term with factor
$\tau^{\mu\nu\rho\sigma}$ after symmetrization, there is an extra
numerical factor $\frac{2}{(m+2)(m+1)}$. Taking all these into account
and interchanging indices $j$ and $m$ in the summation,  we
obtain \begin{equation}
\DD^{\mu\nu\rho\sigma} =
\tau^{\mu\nu\rho\sigma} \sum_{j=0}^\infty
\sum_{m=2j}^{\infty}\frac{2}{(m+2)(m+1)}\pmatrix{2j+2 \cr m+2}(-\epsilon)^m
\left(-\frac{1}{3}\right)^j \LL_{\{j , m-2j \}}
\label{Ljm} \end{equation}
where \begin{equation}
\LL_{\{j , m \}} =
h^{\mu_1\mu_2}\cdots h^{\mu_{2j-1}\mu_{2j}}
u^{\mu_{2j+1}}\cdots u^{\mu_{m+2j}}
\partial_{\mu_{1}\cdots \mu_{m+2j}} \; .
\label{DD2}\end{equation}
When we replace the index $m$ by $m+2j$ (\ref{Ljm}) becomes
\begin{equation}
\DD^{\mu\nu\rho\sigma}  =
\tau^{\mu\nu\rho\sigma} \sum_{j=0}^\infty
\sum_{m=0}^{\infty}\frac{2}{(m+2)(m+1)}\pmatrix{2j+2 \cr
m+2j+2}(-\epsilon)^{m+2j}
\left(-\frac{1}{3}\right)^j
\LL_{\{j , m \}}  \ . \label{m+2j} \end{equation}

We next want to simplify the
differential operator $\LL$ by approximating
it with \begin{equation}
{\cal L}_{\{j , m \}}
=h^{\mu_1\mu_2}\partial_{\mu_{1}\mu_{2}}
\cdots h^{\mu_{2j-1}\mu_{2j}}\partial_{\mu_{2j-
1}\mu_{2j}}u^{\mu_{2j+1}}\partial_{\mu_{2j+1}}\cdots
u^{\mu_{m+2j}}\partial_{\mu_{m+2j}} \ . \label{calL} \end{equation}
Direct evaluation shows that
$ \LL_{\{ 0 , 0 \}} = {\cal L}_{\{0 , 0 \}}\, ,$
$ \LL_{\{1 , 0 \}} = {\cal L}_{\{1 , 0 \}}$ and
$ \LL_{\{0 , 1 \}} = {\cal L}_{\{0 , 1 \}}\, ,$ and this means that
we would be making approximations only in the higher order terms in
(\ref{m+2j}) in replacing $\LL$ by ${\cal L}$. That is, this replacement
leads to the neglect of terms of relative order $\epsilon^2$ in only
the higher terms, as we detail in Appendix \ref{neglect}.

The advantage in making the approximate replacement of $\LL$ is that we
may write \begin{equation}
{\cal L}_{\{j , m \}} = (h^{\alpha\beta}
\partial_{\alpha \beta})^j(u^\alpha\partial_{\alpha})^{m} \; ;
\label{commu}  \end{equation}
this lends itself to the rational approximations that we wish to develop
for the viscous smoothing operator, which we may now write as
\begin{equation}
\DD^{\mu\nu\rho\sigma} =\tau^{\mu\nu\rho\sigma}
 \sum_{j=0}^\infty
\left(-\frac{\epsilon^2}{3}h^{\alpha\beta}\partial
_{\alpha\beta}\right)^j
\sum_{m=0}^{\infty}\frac{2}{(m+2)(m+1)}\pmatrix{2j+2 \cr m+2j+2}(-\epsilon
u^{\alpha}\partial_{\alpha})^m
 \; . \label{DD4}\end{equation}

\section{Rational Approximation for $\DD^{\mu\nu\rho\sigma}$}

An asymptotic expansion such as we have developed can be made
more useful by the technique of resummation \citep{bak75, ben78}, of
which there are various forms.  The general idea is to replace the series
by a rational function of operators in one of several well documented
ways.  For a given function $g(\epsilon)$ with series representation
\begin{equation}
g(\epsilon)=\sum_{i=0}^\infty \gamma_i \epsilon^i
\label{g}
\end{equation}
the Pad\'e approximation of order $[m,n]$ is written as the rational
function \begin{equation}
\Re_{[m,n]}[g(\epsilon)]\equiv \frac{P_m(\epsilon)}{Q_n(\epsilon)}
\label{7.1}
\end{equation}
where $P_m(\epsilon)$ and $Q_n(\epsilon)$ are polynomials in $\epsilon$
of degrees $m$ and $n$, respectively.  The coefficients in these
polynomials are to be fixed according to the condition that the first
$m+n$ coefficients of the expansion of  $\Re_{[m,n]}$ should match the
first $m+n$ coefficients in the series for $g(\epsilon)$.

This procedure, for all its wide acceptance, has not been rigorously
justified, but a considerable experience shows that most of the reasonable
choices are qualitatively better than a truncated asymptotic expansion.
An intuitive explanation is that the rational function can have poles
at the same points that the function to be approximated does.  Because of
this, the range of validity of the expansion may be extended.

In a simple version of resummation, we might try to represent only
a few terms in $\DD^{\mu\nu\rho\sigma}$, say out to order
$\epsilon^2$, as a rational fraction of operators.  The first few terms
from (\ref{DD4}) are \begin{equation}
\DD^{\mu\nu\rho\sigma}=\tau^{\mu\nu\rho\sigma}\left[1-\epsilon
 u^{\alpha}\partial_{\alpha}+(\epsilon u^{\alpha}
\partial_{\alpha})^2
-\frac{\epsilon^2}{3}h^{\alpha\beta}\partial_{\alpha\beta}
+\cdots\right] . \label{128}  \end{equation}
The $[0,1]$ order Pad\'e approximation is
\begin{equation}
\DD^{\mu\nu\rho\sigma}=\frac{\tau^{\mu\nu\rho\sigma}}
{1+\epsilon u^{\alpha}\partial_{\alpha}}
\end{equation}
while the order $[0,2]$ approximation can be expressed as
\begin{equation}
\DD^{\mu\nu\rho\sigma}=\frac{\tau^{\mu\nu\rho\sigma}}
{[1+\epsilon u^{\alpha}\partial_{\alpha}+\frac{1}{3}\epsilon^2h^
{\kappa\lambda}\partial_{\kappa\lambda}]}    \; .
\label{oper} \end{equation}

These representations of the viscous stress tensor should already
provide a considerable improvement on the standard diffusive form,
but we would like to suggest what we think is yet a better version.
In the earlier work on the thermal smoothing problem, it was possible
to obtain an operator that had very accurate limits at large and small
$\epsilon$.  In the normal case, where the transition between the two
situations is smooth, this meant that the approximation would be
reasonably good for all values of $\epsilon$.  We would like to achieve a
similar accuracy for viscous smoothing and, since (\ref{oper}) is always
good for vanishing $\epsilon$, we need to focus on large $\epsilon$ in our
determination.  For this, we propose the following generalization of the
prescription for fixing the numerical coefficients in the rational
approximation.

We introduce a new rational approximation \begin{equation}
\Re^{(N)}_{[m,n]}[g(\epsilon)]\equiv
\frac{P_m(\epsilon)}{Q_n(\epsilon)}
\label{7.2} \end{equation}
where, as before, $P_m(\epsilon)$ and $Q_n(\epsilon)$ are polynomials in
$\epsilon$ whose orders are to be specified at our convenience.  However,
this time, the coefficients are to be chosen according to a more general
matching prescription.  We suppose that the asymptotic series for
$\Re^{(N)}_{[m,n]}$ is  \begin{equation}
\Re^{(N)}_{[m,n]}[g(\epsilon)] = \sum_{i=0}^\infty R_i^{(N)} \epsilon^i \;
\label{Ri} \end{equation}
and we determine its coefficients by the conditions that
$R_i^{(N)} = \gamma_i$ for $i=0,1,,...,m-1$ and for
$i=N+m,..., N+n+m-1$ where the orders $m$ and $n$ are to be specified
as before.  Thus we leave a gap between the highest and lowest order
terms that we choose to match with the asymptotic series of the original
function.

By varying $N$ we control how much influence the higher terms in the
series can have on our rational approximation.   On the other hand,
nothing has been lost since it is clear that when $N=0$ this procedure
reduces to the standard one: $\Re^{(0)}_{[m,n]}=\Re_{[m,n]}$.
However, we shall be especially interested in \begin{equation}
R^{(\infty)}_{[m,n]}[g(\epsilon)]=\lim\limits_{N\rightarrow\infty}
R^{(N)}_{[m,n]}[g(\epsilon)] \; .  \label{7.3} \end{equation}
For example, if we represent $g(\epsilon)$ at the level of
approximation where we have \begin{equation}
\Re^{(N)}_{[0,1]}[g(\epsilon)]
=\frac{\gamma_0}{1-q_N\epsilon} \; , \label{A1} \end{equation}
we obtain
\begin{equation}
q_N=\left(\frac{\gamma_N}{\gamma_0}\right)^{1/N}
\label{A2} \end{equation}
and  \begin{equation}
q_N = {\gamma_{N+1} \over \gamma_N} \; . \end{equation}
We have then, in particular, that
\begin{equation}
\Re^{(\infty)}_{[0,1]}[g(\epsilon)]=
\frac{\gamma_0}{1-q_{\infty}\epsilon}
\label{A3}
\end{equation}
with
\begin{equation}
q_{\infty}=\lim\limits_{N\rightarrow\infty}
\left(\frac{\gamma_N}{\gamma_0}\right)^{1/N} \ . \end{equation}
Since $\lim\limits_{N\rightarrow\infty}(\gamma_0)^{1/N}=1$ for any finite
$\gamma_0$, we have \begin{equation}
q_{\infty} =\lim\limits_{N\rightarrow\infty}(\gamma_N)^{1/N} =
\lim\limits_{N\rightarrow\infty}\frac{\gamma_{N+1}}{\gamma_N}
\; .  \label{A4} \end{equation}

For the present study, we shall use only the representation for the
viscous smoothing operator with $N=\infty$ and $[m,n]=[0,1]$.
It is therefore convenient to use a simple notation for this case,
namely \begin{equation}
{\cal R}[\cdots]\equiv \Re^{(\infty)}_{[0,1]}[\cdots] .
\label{A0} \end{equation}

Returning now to the study of the viscous smoothing operator in
(\ref{DD4}), we apply the resummation procedure to find that
\begin{equation}
{\cal R}\left[\sum_{n=0}^\infty \pmatrix{2j+2 \cr n+2j+2}
(-\epsilon u^{\alpha}\partial_{\alpha})^n\right] =
\frac{1}{1+\epsilon u^{\alpha}\partial_{\alpha}} \ .
\label{DD5}\end{equation}
The disappearance of $j$ from this expression is a result of the large
$N$ limit.

Then we note that \begin{equation}
\sum_{j=0}^\infty \epsilon^{2j}(-\frac{1}{3}h^{\alpha\beta}
\partial_{\alpha\beta})^j=
\frac{1} {1+\frac{1}{3}\epsilon^2h^{\alpha\beta}\partial_{\alpha\beta}}
\; .  \label{add3.2}
\end{equation}
In terms of these newly defined smoothing operators, we may write
\begin{equation}
\DD^{\mu\nu\rho\sigma} =\tau^{\mu\nu\rho\sigma} {\cal D}_x {\cal D}_\tau
 \label{DD6} \end{equation}
where \begin{equation}
{\cal D}_x = \frac{1}
{1+\frac{1}{3}\epsilon^2h^{\alpha\beta}\partial_{\alpha\beta}}\ ,
\qquad \qquad
{\cal D}_\tau = \frac{1}{1+\epsilon u^{\alpha}\partial_{\alpha}}\ .
\end{equation}
Then the viscous stress tensor can be written as
\begin{equation}
\Xi^{\mu\nu} = \frac{16\pi\epsilon}{15} \tau^{\mu\nu\rho\sigma}
{\cal D}_x {\cal D}_\tau (Su_{\rho,\sigma}) \label{add9}
\end{equation}
or, more compactly, within our present level of approximation, as
\begin{equation}
\Xi^{\mu\nu} = {\cal D}_x {\cal D}_\tau  [\mu_0\tau^{\mu\nu\rho\sigma}
u_{\rho,\sigma}]
\label{add10}
\end{equation}
where $\mu_0=\frac{16\pi\epsilon S}{15}$ is the first order shear
viscosity of the Thomas theory.

If we denote the Thomas stress tensor of (\ref{press2}) by
\begin{equation}
\Xi_T^{\mu\nu}=\mu_0\tau^{\mu\nu\rho\sigma}u_{\rho,\sigma}, \end{equation}
then, from (\ref{add10}), (\ref{DD5}) and (\ref{add3.2}), we arrive at
this linear partial differential equation for $\Xi^{\mu\nu}$:
\begin{equation}
(1+\frac{1}{3}\epsilon^2h^{\alpha\beta}\partial_{\alpha\beta})(1+\epsilon
u^{\alpha}\partial_{\alpha})\Xi^{\mu\nu}=\Xi_T^{\mu\nu}  \ .
\label{add11}
\end{equation}
This is an evolution equation or, what meteorologists would call, a
prognostic equation.  However, the operator $u^\mu\partial_\mu$ is
effectively a time derivative and, for times longer than the flight
times of photons, it is negligible in this formula.  So when we are
dealing with nonrelativistic motions, (\ref{add11}) reduces to the
diagnostic equation \begin{equation}
\frac{1}{3} \epsilon^2h^{\alpha\beta}\partial_{\alpha\beta}\Xi^{\mu\nu}
=\Xi_T^{\mu\nu} - \Xi_{\mu\nu} \; . \label{result} \end{equation}
An illustration of an application of this formula is given elsewhere
\citep{che00}.

In the nonrelativistic case, (\ref{result}) could be written as
\begin{equation}
\frac{1}{3} \epsilon^2 \Delta \Xi^{\mu\nu}
=\Xi_T^{\mu\nu} - \Xi^{\mu\nu}   \end{equation}
where $\Delta$ is the three-dimensional Laplacian operator.  However,
this last step involves the neglect of retardation times, which is
not always a safe practice \citep{del72}.

\section{Transport Coefficients}

Our formula for the radiative stress tensor provides a
closure relation that may be used in the standard evolution
equations for the radiative energy density and flux.  These
equations, in conjunction with the fluid dynamical equations,
provide a two-fluid description of radiative fluid dynamics
\citep{sim63, hsi76} that is reasonably
complete.  However, the simpler one-fluid formulation is
more tractable and could be used in certain cases, as when
radiative forces are weak.

In the single fluid approximation we may think of the radiation as
providing a source of dissipation for the fluid.  In that case, we would
treat the fluid as imperfect with appropriate dissipative terms and use
the transfer theory to provide the corresponding transport coefficients.
This version of radiative fluid dynamics has been developed by
\citet{wei72}.  We add to this theory here by providing some resummed
results for the coefficients of viscosity and conductivity.

Since the procedures we have laid down provide formal solutions to
the transfer equation, we may compute the radiative quantities in terms
of the material properties of the radiating medium.  We have described the
calculation of the viscous stress tensor since that is the object needed
to close the hierarchy of moment equations at a low order.  We can
similarly solve for the energy density and radiative flux.

On resumming the expansions for the energy density and flux of the
radiation we obtain expressions for them like that for the viscosity
tensor.  We shall not give the calculations for these other moments
since they parallel those given above and, in any case, our intention in
this section is merely to indicate this other possible direction for
treating the dynamics of radiating fluids.  So we simply report that
\begin{equation}
E =\frac{4\pi}{3} {\cal D}_x {\cal D}_\tau
\left[\left(3-4\epsilon u^{\rho}_{,\rho}\right) S \right]
\label{E} \end{equation}
and \begin{equation}
F^{\mu} = \frac{4\pi\epsilon}{3}
{\cal D}_x {\cal D}_\tau  \left[ h^{\mu\rho}
\left(S_{,\rho}-4Su^{\sigma}u_{\rho,\sigma}\right)\right] \ .
\label{F} \end{equation}
These expressions may be introduced into the radiative stress
tensor, which we write here as \begin{equation}
T^{\mu\nu} = T^{\mu\nu}_0 + \Delta T^{\mu\nu} \end{equation}
where the equilibrium part is given by
\begin{equation}
T_0^{\mu\nu}=4\pi S u^\mu u^\nu - {4\pi\over 3} S h^{\mu\nu} \label{n2}
\end{equation}
and the nonequilibrium contribution is \begin{equation}
\Delta T^{\mu\nu}
= (E-4\pi S) u^\mu u^\nu +F^\mu u^\nu +F^\nu u^\mu
- {1\over 3}(E-4\pi S)h^{\mu\nu} + \Xi^{\mu\nu} \ .  \label{n3}
\end{equation}
With the formula for $E$ just given, we find that \begin{equation}
E - 4\pi S = \frac{-4\pi\epsilon}{3}
{\cal D}_x {\cal D}_\tau (3u^\alpha S_{,\alpha}+4Su_{,\rho}^\rho) \ ,
\label{n4}
\end{equation}
while $F^\mu$ is given by (\ref{F}).

The transport coefficients can be identified from the expression
for entropy generation.  This is given by \citet{wei72} as
\begin{equation}
S^\mu_{,\mu}=\Delta T^{\mu\nu}\left( \frac{u_{\mu,\nu}}{T}-
\frac{u_\mu T_{,\nu}}{T^2}\right) \label{n6} \end{equation}
where $T$ is temperature, the entropy current is \begin{equation}
S^{\mu} = s u^\mu - T^{-1} u_\nu \Delta T^{\mu\nu} \label{entcur}
\end{equation}
and $s$ is the total (matter plus radiation) entropy density in a frame
locally moving with the matter, as advocated by Mihalas (1984).  (We have
made some sign modifications from Weinberg to adapt these expressions to
the metric signature used here.)

The calculations needed in the evaluation of the entropy generation
are not difficult, but they are lengthy and not central to this work, so
we omit them here and merely quote the result, which is \begin{eqnarray}
S^\mu_{,\mu} & = & -\eta h^{\mu\nu}
\left(\frac{T_{,\mu}}{T}-u^\rho
u_{\mu,\rho}\right)\left(\frac{T_{,\nu}}{T}
-u^\rho u_{\nu,\rho}\right)
+\xi\theta^2 \nonumber \\
& + & \frac{2\mu}{T} h^{\mu\rho}h^{\nu\sigma}
(u_{\mu,\nu}+u_{\nu,\mu}-\frac{2}{3}h_{\mu\nu}
\theta)(u_{\rho,\sigma}+
u_{\sigma,\rho}-\frac{2}{3}h_{\rho\sigma}\theta) \label{n16}
\end{eqnarray}
where $\theta=u^\mu_{\; ,\mu}$ and the transport coefficients
are given by
\begin{equation}
\eta={\cal D}_x {\cal D}_\tau
\left(\frac{16\pi\epsilon S}{3T}\right) ,
\label{n19} \end{equation}
\begin{equation}
\xi={\cal D}_x {\cal D}_\tau
\left[\frac{16\pi\epsilon S}{T} \left(\frac{u^\alpha
T_{,\alpha}}{T\theta}+\frac{1}{3}\right)^2\right] \label{n17}
\end{equation}
and
\begin{equation}
\mu={\cal D}_x {\cal D}_\tau  \left(\frac{4\pi}{15}\epsilon S\right).
\label{n18}
\end{equation}
The positivity of the three generalized transport coefficients is
a sufficient condition for the entropy generation rate to be positive.
This is of interest in that a failure of this rate to be positive
would signal a breakdown in the procedures.

The first term in (\ref{n16}) represents entropy generation by thermal
processes --- heat flux, momentum exchange with the radiation field ---
and the corresponding transport coefficient, $\eta$, is a generalized
thermal conductivity.  The second term comes from entropy generation
due to the bulk volume expansion, and its coefficient, $\xi$, is similarly
a form of bulk viscosity. Finally, the third term results from dissipation
by shearing motions, with $\mu$ as a generalization of the Thomas formula.
Each of these transport coefficients take the form of a bare coefficient
(the coefficient obtained in the diffusive limit) renormalized by the
operation of ${\cal D}_x {\cal D}_\tau $.  However, since these formulae
are to be used in connection with a stress tensor of the diffusive form,
these results are applicable only when the flow does not oscillate rapidly
in time or space.

\section{Conclusion}

In the moment method of radiative transfer \citep{tho81, str97}, one
replaces the transfer equation by an infinite hierarchy of moment
equations.  To render this system usable, one truncates it by introducing
a supplementary closure relation among the retained moments.  That
approximation is typically obtained by truncating the expansion for the
intensity at an order corresponding to the level of closure desired.
When we do this at the lowest reasonable level, only the first two moment
equations are retained, and these need to be closed by an expression for
the pressure tensor.  The standard procedure at this level, the Eddington
approximation, fails badly at representing viscous effects in radiative
fluid dynamics \citep{and72}.

Here, we are proposing to keep the first two moment equations since these
contain the main physics of the radiative action on the fluid.  But
rather than following the procedure of truncating the moment hierarchy
with the corresponding Eddington approximation, we
 seek a closure approximation by another route.  Our procedure is to
develop an asymptotic expansion for the intensity in the manner of the
Hilbert expansion of kinetic theory \citep{gra63, caf83}, but we use this
solution only for the purpose of devising an expression for the
viscosity tensor; the detailed solution itself does not play a role in the
theory.  Since the Hilbert expansion is an outer expansion, in the sense
of asymptotic theory \citep{ben78}, we extend its validity by the
technique of resummation so as to render it useful for situations where
the photon mean free path is long

Because of the structure of the transfer equation, our series for the
distribution function turns into a development of $\epsilon {n^\mu}
\partial_{,\mu}$, the operator that appears in the transfer equation
(\ref{retrans}) and that operates iteratively on the source function in
the expansions.  To extract the desired information from this development
we needed to perform integrals over the momentum space of the photons and
this required extensive manipulations.  Since the expansion parameter
$\epsilon$ appears in front of the highest (indeed the only) derivatives
in the problem, we were faced with a problem in singular perturbation
theory, a subject where successful approximation techniques are known but
whose rigor is rarely, if ever, established.

Even with the help of the asymptotic methods, and resummation, the
calculations were demanding and we were led to idealize the problem to the
case with constant mean free path, though this approximation could be
avoided by a suitable choice of optical coordinates, as we noted.  The
resummations, in conjunction with approximations, permitted us to obtain a
compact expression for the radiative viscosity tensor for use at for all
photon free paths.  Our form for this tensor can be written as an integral
operator on the diffusion tensor of \citet{tho30}.  Since this integral
operator is simple, we could see at once how to convert the result into
the linear partial differential equation (\ref{add11}) for $\Xi^{\mu\nu}$.
Depending on whether the fluid dynamics is very relativistic or not, this
equation can be either prognostic or diagnostic but, in any case, it
provides a general closure relation for radiative fluid dynamics.

Asymptotic procedures, when suitably applied, can give very good results,
and they have proved their usefulness in kinetic theory \citep{caf83}.
But of course the results need to be tested, for example, by comparison
with the results of other methods.   To do this, we must solve specific
problems with the methods being examined and to compare results with
those from other methods.  A first test of this approach would be to
compare it with the moment method for this problem since that theory has
been under serious development \citep{tho81, str97} and should
provide a good benchmark for new procedures.  We know already that to
get anything like a reasonable representation of viscous effects in a
moment method, we need to truncate at a higher order than in the standard
Eddington approximation \citep{and72}.

Of course, both these methods give macroscopic descriptions of radiative
fluid dynamics and so cannot satisfy the full boundary and initial
conditions of transport theory, as is known in kinetic theory
\citep{gra63}.  But we suggest that this loss is more than compensated by
a gain in flexibility and usefulness of our macroscopic description.  This
we have seen in applying our method to the problem of entropy generation
in cosmology (to be reported elsewhere).   Those calculations
involve relatively smooth flows and are not as demanding of a method as
more complicated radiating flows such as may arise in the dynamics of
photon bubbles or radiative interaction with vortices and flux tubes in
hot media.  We believe that the study of such more complicated flows
involving strong, possibly turbulent, fluctuations in radiating media
will provide the real test of our approach.

Finally, we would suggest that the methods we have used here could be
useful in other problems where the mean free path of the basic elements
is long compared to scales of variation in the medium.  An example is
turbulent convection where the local mixing length theory encounters
a loss of validity because the mixing length may be comparable to
scales of variation in space and time in the medium.  The results
derived here may point the way to a more useful theory of eddy transport
processes.

\appendix
\section{Useful Angular Integrals \label{angu}}

When we introduce the angular expansions into the integrals for
the moments of the intensity we naturally encounter a variety of
integrals over products of the $l^{\mu}$ and $n^\mu$ alone as well
as over various mixed products of them.  In this appendix we list
some formulas for such products that we have had need of throughout
this work.  Such formulas may be obtained by writing them in terms of
linear combinations of $u^\mu u^\nu$, $\eta^{\mu\nu}$, and so on, and
using special cases to find the coefficients as in previous studies
\citep{and72, tho81, str97, str98}.

First we note that the angular integral of an odd number of factors
$l^\mu$ vanishes.  Therefore, for the general moment over factors
of $l^\mu$ we consider (\ref{ang}) only for even $\ell$. As can be seen,
$M^{\mu_1\dots\mu_2\ell}$ is orthogonal to $u^\mu$ in all suffices and
this simplifies the problem.

The special cases for $\ell=1$ and $\ell=2$ are most commonly encountered
and so it is worth writing them out explicitly: \begin{equation}
\int l^{\mu}l^{\nu}d\Omega  =   -\frac{4\pi}{3}h^{\mu\nu}
\end{equation}
\begin{equation} \int l^\mu l^\nu l^{\rho}l^{\sigma}d\Omega  =
\frac{4\pi}{15}
(h^{\mu\nu} h^{\rho\sigma}+h^{\mu\rho}h^{\nu\sigma}+
h^{\mu\sigma}h^{\nu\rho})   \ .  \label{even} \end{equation}
The general case is found by induction and we have \begin{equation}
M^{\mu_1 \cdots \mu_{2\ell}}  =
(-1)^\ell \frac{4\pi}{2\ell+1}h^{(\mu_1\mu_2} \cdots
h^{\mu_{2\ell-1} \mu_{2\ell})} \; ,  \end{equation}
where the symmetrization is assumed for the indices inside the
parentheses.

We also make use of moments like
\begin{equation}
N^{\mu_1 \mu_2 \cdots \mu_k} = \int
n^{\mu_1}n^{\nu_2} \cdots n^{\mu_k}\; d\Omega \; , \label{Nmu}
\end{equation}
the simplest of which is
\begin{equation}
\int n^{\mu}n^{\nu}d\Omega = \frac{4\pi}{3}(4u^\mu u^\nu-
\eta^{\mu \nu})= \frac{4\pi}{3}(3u^\mu u^\nu-
h^{\mu \nu}). \label{n}
\end{equation}
The general case is easily obtained using the binomial theorem.
On noting that \begin{equation}
n^{\mu_1}\cdots n^{\mu_k} = \sum_{i=0}^{k} \pmatrix{i\cr k\cr}
u^{(\mu_1} \cdots
u^{\mu_i}l^{\mu_{i+1}}\cdots l^{\mu_{k})} \, . \label{bino} \end{equation}
where binomial coefficient is $\pmatrix{i \cr k \cr}=\frac{k!}{i!(k-i)!}$
we have \begin{equation}
N^{\mu_1 \mu_2 \cdots \mu_k} =
\sum_{i=0}^{k} \pmatrix{i \cr k \cr} u^{(\mu_1}
u^{\mu_2} \cdots u^{\mu_{i}} M^{\mu_{i+1} \cdots \mu_{k})} \ .
\label{nInt} \end{equation}

\section{The Operators $\LL$ and ${\cal L}$ \label{neglect}}

In Section \ref{vso} of the text, we have used the following approximation
\begin{equation}
\DD\equiv\sum_{j=0}^{\infty}\sum_{m=0}^{\infty}C_{\{j,m\}}
(-\epsilon)^{m+2j}\LL_{\{j , m \}}
=\sum_{j=0}^{\infty} \sum_{m=0}^{\infty}C_{\{j,m\}}
(-\epsilon)^{m+2j}{\cal L}_{\{j , m \}}
\end{equation}
where $C_{\{j,m\}}$ represents the rest coefficient in operator $\DD$.
To justify this approximation, let us define the difference between
$\LL$ and ${\cal L}$ as
\begin{equation} L_{\{j,m\}}=\LL_{\{j,m\}}-{\cal L}_{\{j,m\}} \ .
\end{equation}
Evidently (as we saw in the text), $L_{\{0,0\}}=L_{\{1,0\}}
=L_{\{0,1\}}=0$.
This leads to
\begin{eqnarray}
\sum_{j=0}^{\infty}\sum_{m=0}^{\infty}(-\epsilon)^{m+2j}
C_{\{j,m\}}L_{\{j,m\}}&=&\sum_{j=2}^{\infty} \sum_{m=0}^{\infty}
(-\epsilon)^{m+2j}C_{\{j,m\}}L_{\{j,m\}} \nonumber \\
&&+\sum_{m=2}^{\infty}(-\epsilon)^{m}C_{\{0,m\}}L_{\{0,m\}} \nonumber \\
&&+\sum_{m=0}^{\infty}(-\epsilon)^{m+2}C_{\{1,m\}}L_{\{1,m\}} .
\label{Osum1}\end{eqnarray}
After shifting indices as we did previously, we get
\begin{eqnarray}
&&\sum_{j=0}^{\infty} \sum_{m=0}^{\infty}(-\epsilon)^{m+2j}
C_{\{j,m\}}L_{\{j,m\}}=\sum_{j=1}^{\infty} \sum_{m=0}^{\infty}
(-\epsilon)^{m+2j}\epsilon^2C_{\{j+1,m\}}L_{\{j+1,m\}}
\nonumber \\
&&+\sum_{m=0}^{\infty}(-\epsilon)^{m}\epsilon^2
(C_{\{0,m+2\}}L_{\{0,m+2\}}+C_{\{1,m\}}L_{\{1,m\}}) .
\label{Osum2}\end{eqnarray}
We can then decompose the summation for
${\cal L}_{\{j,m\}}$ as
\begin{equation}
\sum_{j=0}^{\infty} \sum_{m=0}^{\infty}(-
\epsilon)^{m+2j}C_{\{j,m\}}{\cal L}_{\{j,m\}}
= \sum_{j=1}^{\infty} \sum_{m=0}^{\infty}(-
\epsilon)^{m+2j}C_{\{j,m\}}{\cal L}_{\{j,m\}}+\sum_{m=0}^{\infty}
(-\epsilon)^{m}C_{\{0,m\}}{\cal L}_{\{0,m\}})
. \label{Fsum1}\end{equation}

Thus with the decomposition of (\ref{Osum2})-(\ref{Fsum1})
we have
\begin{eqnarray}
\DD &=&\sum_{j=0}^{\infty} \sum_{m=0}^{\infty}C_{\{j,m\}}
(-\epsilon)^{m+2j}[{\cal L}_{\{j,m\}}+L_{\{j,m\}}]
\nonumber \\ &=&\sum_{j=1}^{\infty}
\sum_{m=0}^{\infty}(-\epsilon)^{m+2j}C_{\{j,m\}}
({\cal L}_{\{j,m\}}+\epsilon^2L_{\{j,m\}})  \\
&+&\sum_{m=0}^{\infty}(-\epsilon)^{m} \left[C_{\{0,m\}}{\cal L}_{\{0,m\}}
+\epsilon^2(C_{\{0,m+2\}}
L_{\{0,m+2\}}+C_{\{1,m\}}L_{\{1,m\}})\right] \nonumber
. \label{DD2.2}\end{eqnarray}

We see that the term by term difference between $\LL$ and
${\cal L}$ is of order $\epsilon^2$.  If we omit in each term
the higher correction associated with $L$, we are left with
\begin{eqnarray}
\DD & = & \sum_{j=1}^{\infty}
\sum_{m=0}^{\infty}(-\epsilon)^{m+2j}C_{\{j,m\}}
{\cal L}_{\{j,m\}}+
\sum_{m=0}^{\infty}(-\epsilon)^{m} C_{\{0,m\}}{\cal L}_{\{0,m\}}\nonumber \\
&=&\sum_{j=0}^{\infty}
\sum_{m=0}^{\infty}(-\epsilon)^{m+2j}C_{\{j,m\}}{\cal L}_{\{j,m\}}
. \label{DD2.3}\end{eqnarray}
This is the approximation we used in the text.


\end{document}